\documentclass[a4paper,12pt]{article}

\usepackage{graphics, epsfig}

\begin{document}

\author{T. Futamase \thanks{E-mail : tof@astr.tohoku.ac.jp},
P. A. Hogan \thanks{E-mail : peter.hogan@ucd.ie} and Y. Itoh
\thanks{E-mail : yousuke@astr.tohoku.ac.jp}\\
\small Astronomical Institute, Graduate School of Science,\\
\small Tohoku University, Sendai 980-8578, Japan\\
}

\title{The Equations of Motion in General Relativity of a Small Charged Black Hole}
\date{PACS numbers: 04.20.-q,04.25.-g,04.25.Nx,04.40.-b}
\maketitle

\begin{abstract}We present the details of a model in General Relativity of a
small charged black hole moving in an external gravitational and
electromagnetic field. The importance of our model lies in the fact that we 
can derive the equations of motion of the black hole from the Einstein--Maxwell 
vacuum field equations \emph{without encountering infinities}. The key assumptions 
which we base our results upon are (a) the black hole is isolated and (b) near the black hole the wave fronts of the radiation generated by its motion are smoothly deformed spheres. The equations of motion which emerge fit the pattern of 
the original De Witt and Brehme equations of motion (after they
 ``renormalise'').  Our calculations 
are carried out in a coordinate system in which the null hypersurface histories of the wave fronts 
can be specified in a simple way, with the result that we obtain a new explicit form, particular to 
our model, for the well--known ``tail term'' in the equations of motion.
\end{abstract}
\thispagestyle{empty}
\newpage

\section{Introduction}\indent
The purpose of this paper is to present a model in
General Relativity of a small charged black hole moving in 
an external gravitational and electromagnetic field. The external
field is a solution of the Einstein--Maxwell field equations and
the history of the black hole is a non--singular time--like
world--line in this background space--time. The presence of the
black hole is envisaged as a small perturbation of the background
space--time which is singular on this world--line ($r=0$). As the
world--line is approached (by letting $r\rightarrow 0$) and the
mass $m$ and charge $e$ of the black hole are taken to be small,
such that in the limit $m\rightarrow 0$ and $e\rightarrow 0$ the
ratios $e/r$ and $m/r$ remain finite, the perturbed gravitational
field (Weyl tensor) is predominantly that of the
Reissner--Nordstrom black hole and the perturbed electromagnetic
field (Maxwell tensor) is predominantly the Coulomb field and the calculations 
described in this paper confirm that these limits can be achieved. These
requirements guide us in our choice of expansions, in integer
powers of $r$, of functions appearing in the metric tensor of the
space--time and the potential 1--form. 

We take the view that the assumed expansions 
restrict the model that we are constructing of a small charged black hole 
moving in external electromagnetic and gravitational fields. Their 
generality is a topic of further study. Assuming that
the moving small black hole is isolated (in particular that there
are no singular null geodesic generators of the histories of the 
wave fronts produced by its motion) we demonstrate how its approximate equations of
motion are derived from the field equations as the requirement that the
wave--fronts emerging from the moving black hole are, in
the neighbourhood of the black hole, smoothly deformed 2--spheres.
Approximations are based solely on the smallness of the mass and
charge of the black hole (in particular slow motion is not
assumed). The equations of motion derived fit the pattern of that
of the original De Witt and Brehme \cite{DB} equations after they have 
``renormalised''. Among the significant features of our approach is (a) the use of a coordinate
system attached to the null hypersurface histories of the
wave--fronts of the radiation produced by the black hole motion,
(b) the emergence of the approximate equations of motion from the
requirement that the wave--fronts are smoothly deformed 2--spheres
near the black hole, (c) the explicitness of the so--called tail term
specific to our model and (d) the absence of infinities arising in our
process.

The topic of this paper has been an active area of research in
general relativity beginning with the De Witt and Brehme \cite{DB}
work (motivated by the classic paper by Dirac \cite{D}). 
The equations of motion have been derived also by Beig
\cite{B} and by Barut and Villarroel \cite{BV}. For important
recent work on the equations of motion with radiation reaction of
small black holes see \cite{M,Q,DW,P}. Some of the latter 
has centered on the problem of
identifying tensor fields in the vicinity of the black hole which
are singular or non--singular, as the case may be, on the
world--line of the small black hole in the background space--time
and the associated ``regularization procedures'' (see, for example
%\cite{Bar,Bar1,Bar2,Bar3,MNS,DMW,Bar4,PfP} 
\cite{Bar}--\cite{PfP} 
and the review \cite{P}). Our
work is complementary to the studies listed above. The precise
relationship to them is a topic of future study however. Nevertheless recent 
work which should particularly be compared with our approach is section 4 in \cite{M}, 
section 19 in \cite{P1} and \cite{GW}, although the 
latter applies only to vacuum background geometries. Earlier
work that has more in common with our approach, but is more
specialized, can be found in for example \cite{HR} (where the
external field is considered weak) and in 
%\cite{NP,RR,HI}
\cite{NP}, \cite{RR} and \cite{HI}
 (where the space--time is less general than here).

The outline of the paper is as follows: In section 2 the Einstein--Maxwell background space--time is 
described in a suitable coordinate system for our purposes and some consequences of the field 
equations required later are derived. The charged black hole space--time is introduced as a perturbation 
of the background space--time in section 3 and the consequences of imposing, approximately, the 
vacuum Einstein--Maxwell field equations are given. In this section the method of extracting the 
equations of motion of the black hole, using the field equations and the properties of the wave--fronts 
near the black hole, is described and the equations of motion are derived. Section 4 is a brief discussion 
highlighting some properties of the equations of motion derived in section 3. To make the paper as self--contained 
as possible calculations required for section 2 are listed in Appendix A while section 3 requires both the 
calculations listed in Appendix A and the more extensive ones listed in Appendix B.

\setcounter{equation}{0}
\section{The Background Space--Time}\indent
We consider a small charged black hole moving in external
gravitational and electromagnetic fields. We model the external
fields by a potential 1--form and a space--time manifold on which
it is defined which are solutions of the vacuum Einstein--Maxwell
field equations. This space--time, which is otherwise unspecified
in this work, contains a time--like world line ($r=0$) on which
the background Maxwell field (the Maxwell tensor field) and the
background gravitational field (the Weyl tensor field) are
non--singular. In the next section the small charged black hole, with mass $m$ and
charge $e$, is introduced as a perturbation of this space--time
which is singular on the world line $r=0$. The perturbed
space--time will be an approximate solution of the vacuum
Einstein--Maxwell field equations having the property that in the
limits $e\rightarrow 0$, $m\rightarrow 0$ and $r\rightarrow 0$
such that the ratios $e/r$ and $m/r$ remain finite the Maxwell
field is dominated by the Coulomb field of the charge and the
gravitational field (Weyl tensor field) is dominated by the
Reissner--Nordstrom field of a charged black hole. To make all of
these requirements more specific we begin by writing  the
line--element of the background in a coordinate system attached to
a family of null hypersurfaces in the space--time having in
general shear and expansion \cite{HT}. The line--element, in terms
of a convenient basis of 1--forms, reads:
\begin{equation}
ds^2=-(\vartheta ^1)^2-(\vartheta ^2)^2+2\,\vartheta ^3\,\vartheta
^4\ ,
\label{eq:metric}
\end{equation}
where 
\begin{eqnarray}
\vartheta ^1&=&r\,p^{-1}({\it e}^{\alpha}\cosh\beta\,dx+{\it
e}^{-\alpha}\sinh\beta\,dy+a\,du)\ ,
\label{eq:tetrad_1}
\\
\vartheta ^2&=&r\,p^{-1}({\it e}^{\alpha}\sinh\beta\,dx+{\it
e}^{-\alpha}\cosh\beta\,dy+b\,du)\ ,
\label{eq:tetrad_2}
\\
\vartheta ^3&=&dr+\frac{c}{2}\,du\ ,
\label{eq:tetrad_3}
\\
\vartheta ^4&=&du\ .
\label{eq:tetrad_4}
\end{eqnarray}
A derivation of this
line--element can be found in \cite{FH}. It is completely general,
containing six functions $p,\ \alpha ,\ \beta ,\ a,\ b,\ c$ of the
four coordinates $x, y, r, u$. It is therefore equivalent to
line--elements constructed by Sachs \cite{S} and by Newman and
Unti \cite{NU}. The particular form we consider here was designed
to allow the Robinson--Trautman \cite{RT1}\cite{RT2}
line--elements to emerge as a convenient special case (this case
corresponds to putting $\alpha =\beta =0$) and this is also a
reason why the form is useful for the subject of the present
paper. The hypersurfaces $u={\rm constant}$ are null and are
generated by the null geodesic integral curves of the vector field
$\partial /\partial r$. The coordinate $r$ is an affine parameter
along these curves and these null geodesics have (complex) shear
\begin{equation}
\label{eq:complex_shear}
\sigma =\frac{\partial\alpha}{\partial r}\,\cosh 2\beta
+i\,\frac{\partial\beta}{\partial r}\ ,
\end{equation} 
and (real) expansion
\begin{equation}
\label{eq:real_expansion}
\rho =\frac{\partial}{\partial r}\log (r\,p^{-1})\ .
\end{equation}

We take this ``background'' space--time to be a solution of the
vacuum Einstein--Maxwell field equations. We take the history of
the small charged black hole to be a (non--singular) time--like
world line in this space--time and we take this world--line to
have equation $r=0$.  Assuming the line--element to be regular on
and in the neighbourhood of $r=0$ we expand the six functions
introduced above in powers of $r$ as follows (the choice of
initial terms here is dictated by the classical work of Fermi
\cite{EF} mentioned below):
\begin{eqnarray}
p&=&P_0(1+q_2\,r^2+q_3\,r^3+\dots )\ ,
\label{eq:expansion_metric_function_p}
\\
\alpha &=&\alpha _2\,r^2+\alpha _3\,r^3+\dots\ ,
\label{eq:expansion_metric_function_alpha}
\\
\beta &=&\beta _2\,r^2+\beta _3\,r^3+\dots\ ,
\label{eq:expansion_metric_function_beta}
\\
a&=&a_1\,r+a_2\,r^2+\dots\ ,
\label{eq:expansion_metric_function_a}
\\
b&=&b_1\,r+b_2\,r^2+\dots\ ,
\label{eq:expansion_metric_function_b}
\\
c&=&c_0+c_1\,r+c_2\,r^2+\dots\ .
\label{eq:expansion_metric_function_c}
\end{eqnarray} 
In the coordinates
$(x, y, r, u)$ the potential 1--form of the background
electromagnetic field takes the form
\begin{equation}
A=L\,dx+M\,dy+K\,du\ .
\label{eq:maxwell_potential}
\end{equation}
In the neighbourhood of
the world line $r=0$ we expand the coefficients of the
differentials in (\ref{eq:maxwell_potential}) in positive powers of $r$ in such a
way that the corresponding electromagnetic field is non--singular
on $r=0$ (since this is the \emph{external} field). A study of
the exterior derivative of the 1--form (\ref{eq:maxwell_potential}) reveals that
the appropriate expansions of the coefficients of the
differentials are:
\begin{eqnarray}
L&=&r^2L_2+r^3L_3+\dots\ ,
\label{eq:expansion_maxwell_L}
\\
M&=&r^2M_2+r^3M_3+\dots\ ,
\label{eq:expansion_maxwell_M}
\\
K&=&r\,K_1+r^2K_2+\dots\ .
\label{eq:expansion_maxwell_K}
\end{eqnarray}The functions appearing
here and in 
(\ref{eq:expansion_metric_function_p})--(\ref{eq:expansion_metric_function_c}) 
as coefficients of the different powers
of $r$ are real--valued functions of $x, y, u$ only. We take the
coordinates $x, y, u$ in the range $(-\infty , +\infty )$ and $r$
in the range $[0, +\infty )$. Had we included in (\ref{eq:maxwell_potential}) 
a term $W\,dr$ we would have had to take $W=W_1r+W_2r^2+\dots\ $, and 
this can be removed by adding a gauge term to (\ref{eq:maxwell_potential}) without changing 
the form of the expansions (\ref{eq:expansion_maxwell_L})--(\ref{eq:expansion_maxwell_K}).

It is easily seen from (\ref{eq:complex_shear}) and (\ref{eq:real_expansion}) that the
complex shear and expansion of the integral curves of $\partial
/\partial r$ are now given respectively by
\begin{eqnarray}
\sigma &=&2\,(\alpha _2+i\beta _2)\,r+3\,(\alpha _3+i\beta
_3)\,r^2+\dots\ ,
\label{eq:expansion_sigma}
\\
\rho &=&\frac{1}{r}-2\,q_2\,r-3\,q_3\,r^2+\dots\ .
\label{eq:expansion_rho}
\end{eqnarray}
Thus near $r=0$ (for small values of $r$) the null hypersurfaces
$u={\rm constant}$ resemble future null cones with vertices on the
world line $r=0$. Following the classical work of Fermi \cite{EF}
(see also \cite{OR1,OR2}) we can, without loss of
generality, take the metric tensor in the neighbourhood of the
world line $r=0$ to be the Minkowskian metric tensor, when convenient in
rectangular Cartesian coordinates and time, up to terms of order
$r^2$. This has led us to the starting terms chosen in
(\ref{eq:expansion_metric_function_p})--(\ref{eq:expansion_metric_function_c}). 
In addition in 
(\ref{eq:expansion_metric_function_p}) and
(\ref{eq:expansion_metric_function_c})  
we can write (see,
for example \cite{NU} and Appendix A)
\begin{equation}
P_0=x\,v^1+y\,v^2+\left\{1-\frac{1}{4}(x^2+y^2)\right\}\,v^3+
\left\{1+\frac{1}{4}(x^2+y^2)\right\}\,v^4\ ,
\label{eq:P_0_in_xyru_coordinate}
\end{equation}
and
\begin{equation}
c_0=1=\Delta \,\log P_0\ ,\qquad \Delta =P_0^2\left
(\frac{\partial ^2}{\partial x^2}+\frac{\partial ^2}{\partial
y^2}\right )\ ,\qquad c_1=-2\,h_0\ ,
\label{eq:c0_c1}
\end{equation} with
\begin{eqnarray}
h_0&=&\eta _{ij}\,a^i\,k^j=\frac{\partial}{\partial u}(\log P_0)\
,
\label{eq:h0}
\\
-P_0\,k^i&=&x\delta ^i_1+y\delta ^i_2+\left
\{1-\frac{1}{4}(x^2+y^2)\right\}\delta ^i_3-\left
\{1+\frac{1}{4}(x^2+y^2)\right\}\delta ^i_4\
.\nonumber
\label{eq:ki}
\\&&
\end{eqnarray}
In these formulas $v^i(u)$ is the
4--velocity of the particle with world line $r=0$ calculated in
rectangular Cartesian coordinates and time $(X^i)$ (Latin indices
take values 1, 2, 3, 4) and $u$ is arc length or proper--time
along this line. $\eta _{ij}={\rm diag}(-1, -1, -1, +1)$ are the
components of the Minkowskian metric tensor in coordinates $(X^i)$
and thus $\eta _{ij}\,v^i\,v^j=+1$. Also $a^i=dv^i/du$ is the
4--acceleration of the particle with world--line $r=0$ (hence
$\eta _{ij}\,v^i\,a^j=0$) and on $r=0$ we have $k^i\,\partial
/\partial X^i=\partial /\partial r$ with $\eta _{ij}k^i\,v^j=1$
(this latter equation applied to (\ref{eq:ki}) yields
(\ref{eq:P_0_in_xyru_coordinate})). 
The operator $\Delta$ in (\ref{eq:c0_c1}) is the Laplacian on the
unit 2--sphere.

The relationship between the rectangular Cartesian coordinates
and time $(X^i)$ and the coordinates $(x, y, r, u)$ near the world
line $r=0$ is given by (see, for example, \cite{NU1})
\begin{equation}
X^i=x^i(u)+r\,k^i\ ,
\label{eq:Xi_to_xi}
\end{equation}
neglecting $O(r^2)$--terms. Thus
the world line $r=0$ has parametric equations $X^i=x^i(u)$ and the
unit tangent vector or 4--velocity vector introduced above has
components $v^i=dx^i/du$. We list in Appendix A some well known
useful formulas arising from (\ref{eq:Xi_to_xi}) which we will refer to
in the sequel.

The electromagnetic field calculated with the 1--form (\ref{eq:maxwell_potential})
has the form
\begin{equation}
F=dA=\frac{1}{2}F_{ab}\,\vartheta ^a\wedge\vartheta ^b\
\label{eq:maxwell_field}
,\end{equation}
with $F_{ab}=-F_{ba}$ and $\vartheta ^a$ given by
(\ref{eq:tetrad_1})--(\ref{eq:tetrad_4}).
Imposing Maxwell's
vacuum field equations $d{}^*F=0$, where the star denotes the
Hodge dual, we find in leading order (in powers of the coordinate
$r$) the following equations to be satisfied by the functions
$L_2,\ M_2,\ K_1$:
\begin{equation}
K_1=P_0^2\left (\frac{\partial L_2}{\partial x}+\frac{\partial
M_2}{\partial y}\right )\ ,\qquad \Delta K_1+2\,P_0^2\left (\frac{\partial L_2}{\partial
x}+\frac{\partial M_2}{\partial y}\right )=0\
\label{eq:equations_for_K1}
,\end{equation}
and
\begin{equation}
\label{eq:equations_for_K1_II}
\frac{\partial K_1}{\partial x}+2\,L_2-\frac{\partial}{\partial
y}\left\{P_0^2\left (\frac{\partial M_2}{\partial
x}-\frac{\partial L_2}{\partial y}\right )\right\}=0\
,\end{equation}
\begin{equation}
\label{eq:equations_for_K1_III}
\frac{\partial K_1}{\partial y}+2\,M_2+\frac{\partial}{\partial
x}\left\{P_0^2\left (\frac{\partial M_2}{\partial
x}-\frac{\partial L_2}{\partial y}\right )\right\}=0\
.\end{equation}
We see immediately that the second of (\ref{eq:equations_for_K1}) is a
consequence of (\ref{eq:equations_for_K1_II}) and
(\ref{eq:equations_for_K1_III}). Also (\ref{eq:equations_for_K1})
implies that $K_1$ is an $l=1$ spherical harmonic:
\begin{equation}
\label{eq:equations_for_K1_IV}
\Delta K_1+2\,K_1=0\ .
\end{equation}
For our purposes the only functions of $x, y, u$ 
appearing in 
(\ref{eq:expansion_maxwell_L})--(\ref{eq:expansion_maxwell_K}) 
that we will require in the sequel are 
$L_2, M_2$ and $K_1$ and so we need not pursue Maxwell`s vacuum field 
equations to higher order in powers of $r$.

The leading terms in the tetrad components of the Maxwell field $F_{ab}$ that we will 
require are given by
\begin{equation}
F_{13}=-2\,P_0\,L_2+O(r)\ ,F_{23}=-2\,P_0\,M_2+O(r)\ ,
F_{34}=K_1+O(r)\ .
\label{eq:F13_F23_F34}
\end{equation}
On $r=0$ we can replace the
basis 1--forms 
(\ref{eq:tetrad_1})--(\ref{eq:tetrad_4})  
by their Minkowskian counterparts
(\ref{A9})--(\ref{A12}) and the coordinates $x, y, r, u$ by the coordinates
$(X^i)$. Thus evaluating (\ref{eq:F13_F23_F34}) on $r=0$ yields
\begin{equation}
\label{eq:L2_M2_K1}
L_2=\frac{1}{2}F_{ij}(u)\,k^i\,\frac{\partial k^j}{\partial x}\ ,\qquad 
M_2=\frac{1}{2}F_{ij}(u)\,k^i\,\frac{\partial k^j}{\partial y}\ ,\qquad 
K_1=F_{ij}\,k^i\,v^j\ .
\end{equation}
where $F_{ij}(u)=-F_{ji}(u)$ are 
the components of the (external) Maxwell field calculated in the coordinates 
$(X^i)$ on $r=0$. We can now verify directly that these expressions for $L_2,\ M_2,\ K_1$ 
satisfy the Maxwell equations (\ref{eq:equations_for_K1})--(\ref{eq:equations_for_K1_IV})
above. For example on substituting $L_2, M_2$ from (\ref{eq:L2_M2_K1}) into 
first of (\ref{eq:equations_for_K1}) yields
\begin{equation}
\label{2.23}
K_1=P_0^2\left (\frac{\partial L_2}{\partial x}+\frac{\partial
M_2}{\partial y}\right )=\frac{1}{2}F_{ij}(u)k^i\,\Delta
k^j=F_{ij}(u)\,k^i\,v^j\ ,
\end{equation}
with the last equality
following from (\ref{A13}) and (\ref{A14}) added together. 

Taking the electromagnetic field above as source, Einstein's field
equations for the background space--time read
\begin{equation}
\label{2.24}
R_{ab}=2\,E_{ab}\ ,
\end{equation} 
where $R_{ab}$ are the
components of the Ricci tensor calculated on the tetrad given via
the 1--forms 
(\ref{eq:tetrad_1})--(\ref{eq:tetrad_4}).  
$E_{ab}$ are the tetrad components of
the electromagnetic energy--momentum tensor calculated using the
Maxwell tensor (\ref{eq:maxwell_field}) according to the formula
\begin{equation}
\label{2.25}
E_{ab}=F_{ca}\,F^{c}{}_b-\frac{1}{4}\,g_{ab}\,F_{cd}\,F^{cd}\ .
\end{equation}
Here $g_{ab}$ are the tetrad components of the
metric tensor and indices on $F_{ab}$ are raised using its
inverse. We will give here only the consequences of the field equations (\ref{2.24}) 
near $r=0$ which will be useful later. To satisfy $R_{33}=2\,E_{33}
+O(r)$ we must have
\begin{equation}
\label{2.26}
q_2=\frac{2}{3}\,P_0^2(L_2^2+M_2^2)\ ,
\end{equation}for $q_2$ appearing in (\ref{eq:expansion_metric_function_p}). 
With $L_2, M_2$ given by (\ref{eq:L2_M2_K1}), and using (\ref{A6}) this can be written
\begin{equation}
%\label{2.27}
q_2=-\frac{1}{6}\,F^p{}_i\,F_{pj}\,k^i\,k^j\ .
\label{eq:q2}
\end{equation}
Here $F_{pj}=F_{pj}(u)$ (and $F^p{}_i=\eta ^{pq}
F_{qi}(u)$) is the Maxwell tensor, in coordinates $(X^i)$ evaluated on $r=0$.
Now to have
\begin{equation}
\label{2.28}
R_{12}=2\,E_{12}+O(r),\ R_{11}-R_{22}=2\,(E_{11}-E_{22})+O(r)\ ,
\end{equation}
requires
\begin{equation}
\label{2.29}
2\,(\alpha _2+i\beta _2)=-\frac{\partial}{\partial\bar\zeta}\left
(a_1+ib_1+4\,P_0^2\frac{\partial q_2}{\partial\bar\zeta}\right )\
,\end{equation}
where $\zeta =x+iy$ 
and a bar will denote complex conjugation. 
Next, for $A=1, 2$,
\begin{equation}
\label{2.30}
R_{A3}=2\,E_{A3}+O(r)\ ,
\end{equation}
provided
\begin{equation}
\label{2.31}
a_1+ib_1+4\,P_0^2\frac{\partial q_2}{\partial\bar\zeta}=2\,P_0^4\frac{\partial}{\partial\zeta}
(P_0^{-2}(\alpha _2+i\beta _2))\ .
\end{equation}
Putting (\ref{2.29}) and (\ref{2.31})
together we arrive at
\begin{equation}
\label{2.32}
\frac{\partial}{\partial\bar\zeta}\left\{P_0^4\frac{\partial}{\partial\zeta}(P_0^{-2}
(\alpha _2+i\beta _2))\right\}=-(\alpha _2+i\beta _2)\ .
\end{equation} 
The approximate field 
equations $R_{AA}=2\,E_{AA}+O(r)$ and $R_{34}=2\,E_{34}+O(r)$ both yield the same equation for the 
function $c_2$ appearing in (\ref{eq:expansion_metric_function_c}) which will not be used in the sequel. The remaining field 
equations in the approximate form
\begin{equation}
\label{2.33}
R_{14}+iR_{24}=2\,(E_{14}+iE_{24})+O(r),\ R_{44}=2\,E_{44}+O(r)\ ,
\end{equation}
are automatically satisfied. 
To check this requires some lengthy calculations. With the Einstein--Maxwell vacuum field equations satisfied in the neighbourhood of the world--line $r=0$ we find the following tetrad components $C_{abcd}$ 
of the Weyl conformal curvature tensor in the neighbourhood of $r=0$:
\begin{eqnarray}
%\label{2.34}
C_{1313}+iC_{1323}&=&6\,(\alpha _2+i\beta _2)+O(r)\ ,
\label{eq:C1313_C1323}
\\
C_{3431}+iC_{3432}&=&\frac{3}{2}P_0^{-1}\left (a_1+ib_1+4\,P_0^2\frac{\partial q_2}{\partial\bar\zeta}\right )
+O(r)\ .
\label{eq:C3431_C3432}
\end{eqnarray}
Using the 1--forms defined in (\ref{A9})--(\ref{A12}) and $q_2$ given by (\ref{eq:q2}) we 
deduce from these that
\begin{eqnarray}
%\label{2.35}
\alpha _2&=&\frac{1}{6}P_0^2C_{ijkl}\,k^i\,\frac{\partial k^j}{\partial x}\,k^k\,\frac{\partial k^l}{\partial x}\ 
,
\label{eq:alpha2}
\\
\beta _2&=&\frac{1}{6}P_0^2C_{ijkl}\,k^i\,\frac{\partial k^j}{\partial x}\,k^k\,\frac{\partial k^l}{\partial y}\ 
,
\label{eq:beta2}
\\
a_1&=&\frac{2}{3}P_0^2\left (C_{ijkl}\,k^i\,v^j\,k^k\,\frac{\partial
		       k^l}{\partial
		       x}+F^p{}_i\,F_{pj}\,k^i\,\frac{\partial
		       k^j}{\partial x}\right )\ ,
\label{eq:a1}
\\   
b_1&=&\frac{2}{3}P_0^2\left (C_{ijkl}\,k^i\,v^j\,k^k\,\frac{\partial
		       k^l}{\partial
		       y}+F^p{}_i\,F_{pj}\,k^i\,\frac{\partial
		       k^j}{\partial y}\right )\ .
\label{eq:b1}
\end{eqnarray}
Here $C_{ijkl}(u)$ are the components of the Weyl tensor of this ``background'' space--time calculated on $r=0$ in 
the coordinates $(X^i)$. Hence they satisfy $C_{ijkl}=-C_{jikl}=-C_{ijlk}=C_{klij}$ and $\eta ^{il}\,C_{ijkl}=0$. Among 
the coefficients of the powers of $r$ in 
(\ref{eq:expansion_metric_function_p})--(\ref{eq:expansion_metric_function_c}) 
we shall only require here the functions $q_2, \alpha _2, 
\beta _2, a_1, b_1$ given by (\ref{eq:q2}) and
(\ref{eq:alpha2})--(\ref{eq:b1}), along with $P_0, c_0$ and $c_1$ 
given in (\ref{eq:P_0_in_xyru_coordinate}) and (\ref{eq:c0_c1}). Using the derivatives of $k^i$ listed in
(\ref{A13})--(\ref{A15}) 
one can directly verify that the field 
equations (\ref{2.29}) and (\ref{2.31}) (and thus (\ref{2.32})) are satisfied by (\ref{eq:q2}) and (\ref{eq:alpha2})--(\ref{eq:b1}).

\setcounter{equation}{0}
\section{Black Hole Perturbation of Background}\indent 
We introduce the small charged black hole as a perturbation of the background Einstein--Maxwell 
space--time described above. With the mass $m$ and charge $e$ considered small of first order we 
write $m=O_1$ and $e=O_1$. The perturbation is introduced by modifying
the expansions 
(\ref{eq:expansion_metric_function_p})--(\ref{eq:expansion_metric_function_c}) as follows:
\begin{eqnarray}
p&=&\hat P_0(1+\hat q_2\,r^2+\hat q_3\,r^3\dots\ )\ ,
\label{3.1.1}
\\
\alpha &=&\hat\alpha _2\,r^2+\hat\alpha _3\,r^3+\dots\ \,
\label{3.1.2}
\\
\beta &=&\hat\beta _2\,r^2+\hat\beta _3\,r^3+\dots\ \,
\label{3.1.3}
\\
a&=&\frac{\hat a_{-1}}{r}+\hat a_0+\hat a_1\,r+\hat a_2\,r^2+\dots\ ,
\label{3.1.4}
\\
b&=&\frac{\hat b_{-1}}{r}+\hat b_0+\hat b_1\,r+\hat b_2\,r^2+\dots\ ,
\label{3.1.5}
\\
c&=&\frac{e^2}{r^2}-\frac{2\,(m+2\,\hat f_{-1})}{r}+\hat c_0+\hat
 c_1\,r+\hat c_2\,r^2\dots.
\label{3.1.6}
\end{eqnarray}
The coefficients of the powers of $r$ are functions of $x, y, u$. The hats are introduced to distinguish these 
coefficients from the background coefficients. Functions here that have non--zero background values are 
assumed to differ from their background values by $O_1$--terms. Functions which vanish in the background 
are assumed to have orders of magnitude: $\hat a_{-1}=O_1,\ \hat a_0=O_1,\ \hat b_{-1}=O_1,\ \hat b_0=
O_1,\ \hat f_{-1}=O_2$. The perturbed potential 1--form is taken to have the form (\ref{eq:maxwell_potential}) with the expansions 
(\ref{eq:expansion_maxwell_L})--(\ref{eq:expansion_maxwell_K}) 
modified to read:
\begin{eqnarray}
L&=&\hat L_0+r^2\hat L_2+r^3\hat L_3+\dots\ ,
\label{3.2.1}
\\
M&=&\hat M_0+r^2\hat M_2+r^3\hat M_3+\dots\ ,
\label{3.2.2}
\\
K&=&\frac{(e+\hat K_{-1})}{r}+\hat K_0+r\,\hat K_1+r^2\hat K_2+\dots\
.
\label{3.2.3}
\end{eqnarray}
Again functions here 
which are non--zero in the background are assumed to differ from their background values by $O_1$--terms and 
those functions which vanish in the background are taken to have the orders of magnitude: $\hat L_0=O_1,\ 
\hat M_0=O_1,\ \hat K_{-1}=O_2,\ \hat K_0=O_1$. All of these expansions have been chosen in order to satisfy the Reissner--Nordstrom black hole limit described in the Introduction. A careful study suggests that they are the minimal choice of 
expansions necessary to satisfy the black hole limit. We consider, however, 
that these assumptions restrict the model that we are constructing of a small black hole moving in external 
electromagnetic and gravitational fields and that their generalisation or otherwise is a topic for further study. In order 
to simplify the presentation we make the further assumption that near $r=0$ the potential 1--form is predominantly 
the Li\'enard--Wiechert 1--form (given by $A=e\,(r^{-1}-h_0)\,du$ up to a gauge term). This is achieved with the further specialisation of the functions $\hat L_0, \hat M_0, \hat K_0$ to satisfy
\begin{equation}
\label{3.2a}
\hat L_0=O_2\ ,\ \hat M_0=O_2\ ,\ \hat K_0=-e\,h_0+O_2\
,\end{equation}
with $h_0$ given by (\ref{eq:h0}).

We emphasise that our approach is \emph{not} one of ``matched asymptotic expansions" in which so--called ``near 
zone" and ``far zone" expansions of field variables are matched in an intermediate ``buffer zone". In our 
perturbed space--time each function appearing in the metric tensor and the potential 1--form has only 
one expansion in powers of $r$ given by (3.1)--(3.6) and (3.7)--(3.9) and all of our deductions, 
including the equations of motion, are made using these expansions and the vacuum Einstein--Maxwell field 
equations.

The expansions we have chosen here ensure that the null--hypersurfaces $u={\rm constant}$ in the perturbed 
space--time, which will play the role of the histories of the wave fronts of the radiation produced by 
the black hole motion, are approximately future null--cones for small $r$. Thus the wave fronts can be approximately 
2--spheres near the black hole. Neglecting $O(r^4)$--terms the line--elements induced on these null hypersurfaces 
are given by
\begin{equation}
\label{3.3}
ds_0^2=-r^2\hat P_0^{-2}\,(dx^2+dy^2)\qquad{\rm with}\qquad\hat
P_0=P_0\,(1+Q_1+Q_2+O_3)\ ,
\end{equation}
with $P_0$ given by (\ref{eq:P_0_in_xyru_coordinate}) and $Q_1=O_1,\ Q_2=O_2$. We shall require the perturbations of these 2--spheres, 
described by the functions $Q_1$ and $Q_2$ here, to be smooth so that as functions of $x, y$ they are non--singular 
(in particular at $x=\pm\infty$ and/or $y=\pm\infty$) so that no ``directional singularities'' occur violating the 
notion of an isolated black hole. We will insist that in general in solving the Einstein--Maxwell vacuum
field equations for the perturbed space--time and the perturbed electromagnetic field that such directional singularities are unacceptable. Then it will follow from the Einstein--Maxwell field equations that 
\emph{necessary conditions for the 2--surfaces with line--elements (\ref{3.3}) to be 
smooth, non--trivial deformations of 2--spheres will be the equations of motion of the black hole}. Such an origin of 
equations of motion can be found, for example, in \cite{NP}, \cite{RR} and \cite{HI} (where the space--time 
is less general than here) and in \cite{HR} (where the external field is weak). 

The left hand sides of the vacuum Einstein--Maxwell field equations for the perturbed space--time are now power series in $r$, 
each with a finite number of terms involving inverse powers of $r$ and an infinite number of terms involving positive powers 
of $r$. The coefficients of the leading terms in each case are listed in Appendix B. They are functions of $x, y, u$ and we require 
them to be small (in terms of $e$ and $m$) with just sufficient accuracy
to enable us to derive the functions in the expansions 
(\ref{3.1.1})--(\ref{3.1.6})  
and (\ref{3.2.1})--(\ref{3.2.3}) required for the equations of motion of the black hole with an $O_3$--error (to include electromagnetic radiation reaction
etc.). The errors that we tolerate to achieve this are indicated in Appendix B. Some of these errors could be reduced but with no 
influence on the equations of motion to the accuracy that we calculate here. We begin by calculating the equations of motion 
with an $O_2$--error. In this case we require from Einstein`s equations
(for the moment requiring 
(\ref{B.6.11}) and (\ref{B.6.17}) 
only to be small of second order) 
\begin{equation}
\label{3.4}
\hat a_{-1}=-4\,e\,P_0^2L_2+O_2=O_1\ ,\qquad\hat
b_{-1}=-4\,e\,P_0^2M_2+O_2=O_1\ ,
\end{equation}
and thus from Maxwell`s 
equations (with (\ref{B.4.8}) required at this stage to be small of first order) we have
\begin{equation}
\label{3.5}
\hat K_1=P_0^2\left (\frac{\partial L_2}{\partial x}+\frac{\partial
	       M_2}{\partial y}\right )+O_1=K_1+O_1\ ,
\end{equation}
with $K_1$ given in (\ref{eq:L2_M2_K1}). We also need 
\begin{equation}
\label{3.6}
\hat c_0=1+\Delta Q_1+2\,Q_1+8\,e\,F_{ij}\,k^i\,v^j+O_2\ ,
\end{equation}
which emerges from the requirement that (\ref{B.6.3}) be 
small of second order. When these are substituted into the final Einstein field equation (given by requiring (\ref{B.6.30}) for 
the moment to be small of second order) the 
differential equation that emerges for $Q_1$ is 
\begin{equation}
\label{3.9}
-\frac{1}{2}\Delta (\Delta
Q_1+2\,Q_1)=6\,m\,a_i\,p^i-6\,e\,F_{ij}\,p^i\,v^j+O_2\ .
\end{equation}Here $p^i=h^i_j\,k^j$ with $h^i_j=\delta ^i_j-v^i\,v_j$ is 
the projection of $k^i$ orthogonal to $v^i$ and satisfies $p^i\,p_i=-1, p^i\,v_i=0$ and since $p^i=k^i-v^i$ we see from (\ref{A13}) and (\ref{A14}) that 
$\Delta p^i+2\,p^i=0$ so that each component of $p^i$ is an $l=1$ spherical harmonic (a spherical harmonic $Q$ of order $l$ is a smooth 
solution of $\Delta Q+l(l+1)\,Q=0$). Integrating (\ref{3.9}), discarding the homogeneous solution with directional singularities, we have 
\begin{equation}
\label{3.10}
\Delta Q_1+2\,Q_1=6\,m\,a_i\,p^i-6\,e\,F_{ij}\,p^i\,v^j+A(u)+O_2\ ,
\end{equation}
where $A(u)=O_1$ is arbitrary. The first two terms on the right hand side 
will now lead to directional singularities in $Q_1$ unless $m\,a_i\,p^i-e\,F_{ij}\,p^i\,v^j=O_2$ for all $p^i$, which implies the equations of motion in first 
approximation:
\begin{equation}
\label{3.11}
m\,a_i=e\,F_{ij}\,v^j+O_2\ .
\end{equation}
Hence we see the Lorentz 4--force due to the external electromagnetic field appearing. With (\ref{3.11}) satisfied the 
equation (\ref{3.10}) simplifies and the directional singularity free $Q_1$ is a linear combination of an $l=1$ spherical harmonic (the smooth homogeneous 
solution) and an $l=0$ spherical harmonic (the particular integral $A(u)/2$). In general perturbations 
of (\ref{3.3}) described by $Q_1$ or $Q_2$ which are $l=0$ or $l=1$ spherical harmonics are trivial. When the 2--surfaces with line--element (\ref{3.3}) are 
viewed against the Euclidean 3--space background such terms in $Q_1$ or $Q_2$ merely infinitesimally change the radius of the sphere (if $l=0$) or 
infinitesimally displace the origin of the sphere (if $l=1$) . We shall consistently discard such ``perturbations'' which means that in this case we shall take $Q_1=0$.  

From now on we take $Q_1=0$ in (\ref{3.3}) and our objective is to solve the Einstein--Maxwell vacuum field equations for the perturbed space--time and 
the perturbed Maxwell field with sufficient accuracy to enable us derive the function $Q_2$ in (\ref{3.3}) and the equations of motion with an $O_3$-error. The 
functions of $x, y, u$ appearing in the potential 1--form (as
coefficients in the expansions 
(\ref{3.2.1})--(\ref{3.2.3}) 
that are 
involved are $\hat L_0,\ \hat M_0,\ \hat K_{-1}$ and the $O_1$--part of $\hat K_1$ and $\hat K_0$. Two of Maxwell's 
equations ((\ref{B.4.0}) and (\ref{B.4.3}) small of third order) read
\begin{equation}
\label{3.12}
\frac{\partial\hat K_{-1}}{\partial x}+6\,e^2L_2+\frac{\partial}{\partial y}\left\{P_0^2\left (\frac{\partial\hat M_0}{\partial x}-\frac{\partial\hat L_0}{\partial y}\right )\right\}=
O_3\ ,
\end{equation}
and 
\begin{equation}
\label{3.13}
\frac{\partial\hat K_{-1}}{\partial y}+6\,e^2M_2-\frac{\partial}{\partial x}\left\{P_0^2\left (\frac{\partial\hat M_0}{\partial x}-\frac{\partial\hat L_0}{\partial y}\right )\right\}=
O_3\ .
\end{equation}
These imply 
\begin{equation}
\label{3.14}
\Delta\hat K_{-1}+6\,e^2F_{ij}\,k^i\,v^j=O_3\ ,
\end{equation}
which incidentally is (\ref{B.4.6}), another of the Maxwell equations. This latter equation can immediately be integrated 
without introducing directional singularities to yield
\begin{equation}
\label{3.15}
\hat K_{-1}=3\,e^2F_{ij}\,k^i\,v^j+e^2K(u)+O_3\ ,
\end{equation}
where $K(u)=O_0$ is an arbitrary function. From (\ref{3.12}) and (\ref{3.13}) we obtain
\begin{equation}
\label{3.16}
\Delta\left\{P_0^2\left (\frac{\partial\hat M_0}{\partial x}-\frac{\partial\hat L_0}{\partial y}\right )\right\}-6\,e^2P_0^2F_{ij}\,\frac{\partial k^i}{\partial x}\,
\frac{\partial k^j}{\partial y}=O_3\ .
\end{equation}
Now the second term on the left here is an $l=1$ spherical harmonic. In fact we can write 
\begin{equation}
\label{3.17}
P_0^2F_{ij}\,\frac{\partial k^i}{\partial x}\,\frac{\partial
k^j}{\partial y}=F^*_{ij}\,k^i\,v^j\ ,
\end{equation}
where $F^*_{ij}(u)$ is the dual of $F_{ij}(u)$, on account of 
the relationship
\begin{equation}
\label{3.18}
\epsilon _{pqkl}\,k^p\,\frac{\partial k^q}{\partial
x}=k_k\,\frac{\partial k_l}{\partial y}-k_l\,\frac{\partial
k_k}{\partial y}\ ,
\end{equation}
where $\epsilon _{ijkl}$ is the Levi--Civita permutation symbol in four
dimensions and $k^i$ is given by (\ref{eq:ki}). 
We integrate (\ref{3.16}) to obtain
\begin{equation}
\label{3.19}
P_0^2\left (\frac{\partial\hat M_0}{\partial x}-\frac{\partial\hat
      L_0}{\partial y}\right
      )=-3\,e^2P_0^2F^*_{ij}\,k^i\,v^j+e^2S(u)+O_3\ ,
\end{equation}
with 
$S(u)=O_0$ arbitrary. This is the extent to which we shall require a knowledge of $\hat L_0$ 
and $\hat M_0$. Substituting (\ref{3.15}) and (\ref{3.19}) into the first order equations (\ref{3.12}) and (\ref{3.13}) verifies that they are now satisfied.We now 
look for the $O_1$--parts of $\hat L_2$ and $\hat M_2$. We shall write $\hat L_2=L_2+l_2+O_2$ and $\hat M_2=M_2+m_2+O_2$ with 
$l_2=O_1$ and $m_2=O_1$. We can obtain $l_2$ and $m_2$ indirectly as follows: in the light of (\ref{3.15}) and (\ref{3.19}) two of Einstein's field 
equations yield algebraically ((\ref{B.6.11}) and (\ref{B.6.17}) required to be $O_3$)
\begin{eqnarray}
\hat a_{-1}&=&-4\,e\,P_0^2(L_2+l_2)+6\,e^2P_0^2F^p{}_i\,F_{pj}\,k^i\,\frac{\partial k^j}{\partial x}+4\,e^2P_0^2M_2\,S(u)\nonumber\\
&&-4\,e^2P_0^2L_2\,K(u)+O_3\ ,
\label{3.20.1}
\\
\hat b_{-1}&=&-4\,e\,P_0^2(M_2+m_2)+6\,e^2P_0^2F^p{}_i\,F_{pj}\,k^i\,\frac{\partial k^j}{\partial y}-4\,e^2P_0^2L_2\,S(u)\nonumber\\
&&-4\,e^2P_0^2M_2\,K(u)+O_3\  ,
\label{3.20.2}
\end{eqnarray}
Define two functions $\hat A$ and $\hat B$ by
\begin{equation}
\label{3.21}
\hat a_{-1}+8\,P_0^2L_2\hat K_{-1}=-4\,e\,P_0^2L_2+\hat A\ ,
\end{equation}
and
\begin{equation}
\label{3.22}
\hat b_{-1}+8\,P_0^2M_2\hat K_{-1}=-4\,e\,P_0^2M_2+\hat B\ .
\end{equation}
Now two of Einstein's field equations 
((\ref{B.6.5}) and (\ref{B.6.8}) both $O_3$) can be written as equations for $\hat A$ 
and $\hat B$ neatly in the complex form
\begin{equation}
\label{3.23}
\frac{\partial}{\partial\bar\zeta}(\hat A+i\hat
B)=-8\,e^2P_0^2(L_2+iM_2)^2-2\,e^2(\alpha _2+i\beta _2)+O_3\ .
\end{equation}
Here $L_2, M_2, \alpha _2, \beta _2$ are the background
functions given by (\ref{eq:L2_M2_K1}), (\ref{eq:alpha2}) and (\ref{eq:beta2}). 
We note that $\alpha _2+i\beta _2$ can 
be expressed as a derivative with respect to $\bar\zeta$ given by (\ref{2.32}). Similarly one can show that 
\begin{equation}
\label{3.24}
P_0^2(L_2+iM_2)^2=-2\,\frac{\partial}{\partial\bar\zeta}\left
							 (P_0^4(L_2+iM_2)\,\frac{\partial}{\partial\zeta}(L_2+iM_2)\right )\ .
\end{equation}
Hence (\ref{3.23}) can immediately be integrated giving a solution which is free of directional singularities:
\begin{eqnarray}
\label{3.25}
\hat A+i\hat B&=&2\,e^2P_0^4\frac{\partial}{\partial\zeta}\{P_0^{-2}(\alpha _2+i\beta _2)\}+16\,e^2P_0^4(L_2+iM_2)\,\frac{\partial}
{\partial\zeta}(L_2+iM_2)\nonumber\\
&&+e^2U(u)+ie^2V(u)+O_3\ ,
\end{eqnarray}
where $U(u)=O_0$ and $V(u)=O_0$ are functions of integration. Combining
(\ref{3.20.1})  with (\ref{3.21}) and 
(\ref{3.20.2})  with (\ref{3.22}) and using the solution (\ref{3.25}) means that we have now determined the functions $l_2$ and $m_2$ to be
\begin{eqnarray}
\label{3.26}
l_2&=&e\,F_{pq}\,k^p\,\frac{\partial k^q}{\partial x}\,F_{ij}\,k^i\,v^j+\frac{1}{2}e\,F^p{}_i\,F_{pj}\,k^i\,\frac{\partial k^j}{\partial x}-\frac{1}{6}e\,C_{ijkl}\,k^i\,v^j\,k^k\,\frac{\partial k^l}{\partial x}
\nonumber\\&&+\frac{1}{2}e\,S(u)\,F_{ij}\,k^i\,\frac{\partial k^j}{\partial y}+\frac{1}{2}e\,K(u)\,F_{ij}\,k^i\,\frac{\partial k^j}{\partial x}-\frac{1}{4}e\,P_0^{-2}U(u)+O_2\ ,\nonumber\\&&\\
m_2&=&e\,F_{pq}\,k^p\,\frac{\partial k^q}{\partial y}\,F_{ij}\,k^i\,v^j+\frac{1}{2}e\,F^p{}_i\,F_{pj}\,k^i\,\frac{\partial k^j}{\partial y}-\frac{1}{6}e\,C_{ijkl}\,k^i\,v^j\,k^k\,\frac{\partial k^l}{\partial y}
\nonumber\\&&-\frac{1}{2}e\,S(u)\,F_{ij}\,k^i\,\frac{\partial
k^j}{\partial x}+\frac{1}{2}e\,K(u)\,F_{ij}\,k^i\,\frac{\partial
k^j}{\partial y}-\frac{1}{4}e\,P_0^{-2}V(u)+O_2\ .\nonumber\\
\end{eqnarray}
One of Maxwell's equations ((\ref{B.4.8}) required to be $O_2$) now gives us directly
\begin{equation}
\label{3.27}
\hat K_1=F_{ij}\,k^i\,v^j+P_0^2\left (\frac{\partial l_2}{\partial
				x}+\frac{\partial m_2}{\partial y}\right
				)-14\,e\,q_2+O_2\ ,
\end{equation}
and we note that $q_2$ is given by (\ref{eq:q2}). 

We now turn our attention 
to the functions describing the perturbations of the metric tensor. We require $\hat a_0, \hat b_0, \hat f_{-1}$ and $\hat c_0$. Defining the variables
\begin{equation}
\label{3.29}
\hat {\cal A}=\hat a_0+4\,P_0^2L_2\,\hat K_0\ ,\qquad {\rm and}\qquad
\hat {\cal B}=\hat b_0+4\,P_0^2M_2\,\hat K_0\  ,
\end{equation}
two of Einstein's equations ((\ref{B.6.6}) and (\ref{B.6.9}) taken to be $O_2$) 
can be written as the one complex equation
\begin{equation}
\label{3.30}
\frac{\partial}{\partial\bar\zeta}(\hat {\cal A}+i\hat {\cal
B})=4\,m\,P_0^2(L_2+iM_2)^2+4\,m\,(\alpha _2+i\beta _2)+O_2\ .
\end{equation}
This is integrated in similar fashion to (\ref{3.23}). The solution 
should have the property that $P_0^{-1}\hat a_0$ and $P_0^{-1}\hat b_0$ 
are free of directional singularities since it is in this way that $\hat a_0$ and $\hat b_0$ appear in the 
metric tensor. The solutions of (\ref{3.30}) therefore that we require are
\begin{eqnarray}
\label{3.31}
\hat a_0&=&2\,e\,P_0^2a_p\,k^p\,F_{ij}\,k^i\,\frac{\partial k^j}{\partial x}-2\,m\,P_0^2F^p{}_i\,F_{pj}\,k^i\,\frac{\partial k^j}{\partial x}\nonumber\\
&&-4\,m\,P_0^2
F_{pq}\,k^p\,v^q\,F_{ij}\,k^i\,\frac{\partial k^j}{\partial x}-\frac{4}{3}m\,P_0^2C_{ijkl}\,k^i\,v^j\,k^k\,\frac{\partial k^l}{\partial x}\nonumber\\
&&+X(u)+O_2\ ,\\
\hat b_0&=&2\,e\,P_0^2a_p\,k^p\,F_{ij}\,k^i\,\frac{\partial k^j}{\partial y}-2\,m\,P_0^2F^p{}_i\,F_{pj}\,k^i\,\frac{\partial k^j}{\partial y}\nonumber\\
&&-4\,m\,P_0^2
F_{pq}\,k^p\,v^q\,F_{ij}\,k^i\,\frac{\partial k^j}{\partial y}-\frac{4}{3}m\,P_0^2C_{ijkl}\,k^i\,v^j\,k^k\,\frac{\partial k^l}{\partial y}\nonumber\\
&&+Y(u)+O_2\ ,
\end{eqnarray}
where $X(u)=O_1$ and $Y(u)=O_1$ are functions of integration. We note that 
$\hat a_0$ and $\hat b_0$ possess directional singularities but  $P_0^{-1}\hat a_0$ and $P_0^{-1}\hat b_0$ do not, as required. Another of Einstein's field equations ((B.50) required to be $O_3$) provides us 
with the equation
\begin{equation}
\label{3.32}
\Delta\hat f_{-1}-2\,e^2a_i\,k^i+m\,P_0^2\left (\frac{\partial}{\partial
					  x}(P_0^{-2}\hat
					  a_{-1})+\frac{\partial}{\partial y}(P_0^{-2}\hat b_{-1})\right )=O_3\ ,
\end{equation}
which using (\ref{3.20.1}) and (\ref{3.20.2})  can be rewritten as
\begin{equation}
\label{3.33}
\Delta (\hat f_{-1}-e^2a_i\,k^i-2\,m\,e\,F_{ij}\,k^i\,v^j)=O_3\  ,
\end{equation}
from which we conclude that
\begin{equation}
\label{3.34}
\hat f_{-1}=e^2a_i\,k^i+2\,m\,e\,F_{ij}\,k^i\,v^j+G(u)+O_3\ ,
\end{equation}
with $G(u)=O_2$ a function of integration. Now we obtain directly from an Einstein equation ((\ref{B.6.3}) taken $O_3$) the function 
$\hat c_0$. After some considerable simplification it can be written
\begin{eqnarray}
\label{3.35}
\hat c_0&=&1+8\,e\,F_{ij}\,k^i\,v^j+\Delta Q_2+2\,Q_2-4\,e^2C_{ijkl}\,k^i\,v^j\,k^k\,v^l\nonumber\\
&&+8\,e^2F^{ij}\,F_{ij}+24\,e^2(F_{ij}\,k^i\,v^j)^2-24\,e^2F^p{}_i\,F_{pj}\,k^i\,v^j\nonumber\\
&&+\frac{65}{3}\,e^2F^p{}_i\,F_{pj}\,k^i\,k^j+16\,e^2K(u)\,F_{ij}\,k^i\,v^j+4\,e^2U(u)\,\frac{\partial}{\partial x}(\log P_0)\nonumber\\
&&+4\,e^2V(u)\,\frac{\partial}{\partial y}(\log P_0)+O_3\ .
\end{eqnarray}
The perturbation in the function $\hat c_1=c_1+O_1$ is now obtained from 
(\ref{B.6.4}) or (\ref{B.6.27}) 
required to be $O_2$. We shall not need this function in the sequel. The remaining quantities 
in (\ref{B.4.0})--(\ref{B.4.8})
and 
(\ref{B.6.1})--(\ref{B.6.29}) 
can now be evaluated and their orders of magnitude are indicated in Appendix B. 
The differential equation for $Q_2$ emerges from the remaining quantity 
(\ref{B.6.30}) required to be $O_3$. This expression is 
lengthy and so we split it into a sum of three terms which we call $T_1, T_2, T_3$. The first of these terms involves only $\hat a_{-1}$ and $\hat b_{-1}$ and is given by
\begin{eqnarray}
\label{3.36}
T_1&=&-\left (\frac{\partial\hat a_{-1}}{\partial x}\right
)^2-\left (\frac{\partial\hat b_{-1}}{\partial y}\right
)^2-\frac{1}{2}\,\left (\frac{\partial\hat a_{-1}}{\partial
y}\right )^2-\frac{1}{2}\,\left (\frac{\partial\hat
b_{-1}}{\partial x}\right )^2\nonumber\\
&&-\frac{\partial\hat a_{-1}}{\partial y}\,
\frac{\partial\hat b_{-1}}{\partial x}+4\,\hat b_{-1}\,\frac{\partial\hat b_{-1}}{\partial
y}\,P_0^{-1}\frac{\partial P_0}{\partial y}+4\,\hat
a_{-1}\,\frac{\partial\hat a_{-1}}{\partial
x}\,P_0^{-1}\frac{\partial P_0}{\partial x}\nonumber\\
&&+2\,\hat
a_{-1}\frac{\partial\hat b_{-1}}{\partial
y}\,P_0^{-1}\frac{\partial P_0}{\partial x}+2\,\hat b_{-1}\frac{\partial\hat a_{-1}}{\partial
x}\,P_0^{-1}\frac{\partial P_0}{\partial y}+2\,\hat
a_{-1}\,\frac{\partial\hat b_{-1}}{\partial
x}\,P_0^{-1}\frac{\partial P_0}{\partial y}\nonumber\\
&&+2\,\hat
b_{-1}\,\frac{\partial\hat a_{-1}}{\partial
y}\,P_0^{-1}\frac{\partial P_0}{\partial x}-2\,(\hat a_{-1})^2P_0\,\frac{\partial ^2}{\partial
x^2}(P_0^{-1})-2\,(\hat b_{-1})^2P_0\,\frac{\partial
^2}{\partial y^2}(P_0^{-1})\nonumber\\
&&-8\,\hat a_{-1}\,\hat b_{-1}\,P_0^{-2}\frac{\partial
P_0}{\partial x}\,\frac{\partial P_0}{\partial y}-\hat
a_{-1}\,\left\{\frac{\partial ^2\hat a_{-1}}{\partial
x^2}+\frac{\partial ^2\hat b_{-1}}{\partial x\partial
y}\right\}\nonumber\\
&&-\hat b_{-1}\,\left\{\frac{\partial ^2\hat
b_{-1}}{\partial y^2}+\frac{\partial ^2\hat a_{-1}}{\partial
x\partial y}\right\}\ .
\end{eqnarray}
To evaluate this we only require the leading terms in $\hat a_{-1}$ and $\hat b_{-1}$ 
given in (\ref{3.20.1}) and (\ref{3.20.2}). By (\ref{eq:F13_F23_F34}) these can be written
\begin{equation}
\label{3.37}
\hat a_{-1}=2\,e\,P_0F_{13}+O_2\ ,\qquad {\rm and}\qquad \hat
b_{-1}=2\,e\,P_0F_{23}+O_2\ .
\end{equation} 
Direct calculation reveals
\begin{equation}
\label{3.38}
\frac{\partial\hat a_{-1}}{\partial y}=-\frac{\partial\hat
b_{-1}}{\partial x}+O_2=2e\,\left\{\frac{\partial P_0}{\partial
y}\,F_{13}-\frac{\partial P_0}{\partial
x}\,F_{23}+F_{12}\right\}+O_2\ ,
\end{equation}
\begin{equation}
\label{3.39}
\frac{\partial\hat a_{-1}}{\partial x}=\frac{\partial\hat
b_{-1}}{\partial y}+O_2=2e\,\left\{\frac{\partial P_0}{\partial
x}\,F_{13}+\frac{\partial P_0}{\partial
y}\,F_{23}-F_{34}\right\}+O_2\ ,
\end{equation}
\begin{eqnarray}
\label{3.40}
\frac{\partial ^2\hat a_{-1}}{\partial x\,\partial
y}&=&e\,\left\{-P_0^{-1}+2\,\frac{\partial ^2P_0}{\partial
y^2}-2\,P_0^{-1}\left (\frac{\partial P_0}{\partial x}\right
)^2\right\}\,F_{23}-2\,e\,P_0^{-1}F_{24}\nonumber\\
&&+2\,e\,P_0^{-1}\frac{\partial P_0}{\partial x}\,\frac{\partial
P_0}{\partial y}\,F_{13}+2\,e\,P_0^{-1}\frac{\partial
P_0}{\partial x}\,F_{12}-2\,e\,P_0^{-1}\frac{\partial
P_0}{\partial y}\,F_{34}\ ,\nonumber\\&&\\ \frac{\partial ^2\hat
b_{-1}}{\partial x\,\partial
y}&=&e\,\left\{-P_0^{-1}+2\,\frac{\partial ^2P_0}{\partial
x^2}-2\,P_0^{-1}\left (\frac{\partial P_0}{\partial y}\right
)^2\right\}\,F_{13}-2\,e\,P_0^{-1}F_{14}\nonumber\\
&&+2\,e\,P_0^{-1}\frac{\partial P_0}{\partial x}\,\frac{\partial
P_0}{\partial y}\,F_{23}-2\,e\,P_0^{-1}\frac{\partial
P_0}{\partial y}\,F_{12}-2\,e\,P_0^{-1}\frac{\partial
P_0}{\partial x}\,F_{34}\ ,\nonumber\\&&
\end{eqnarray}
where $F_{ab}$ are the components of the external electromagnetic 
field evaluated on $r=0$ and referred to the tetrad given by (\ref{A9})--(\ref{A12}). Substitution
in (\ref{3.36}) leads to the simplification
\begin{equation}
\label{3.41}
T_1=4\,e^2\left\{(F_{13})^2+(F_{23})^2\right\}+8\,e^2\left\{F_{13}\,F_{14}+F_{23}\,F_{24}\right\}
-8\,e^2(F_{34})^2+O_3\ .
\end{equation}
This is further reduced by
noting that
\begin{equation}
\label{3.42}
(F_{13})^2+(F_{23})^2=-F^p{}_i\,F_{pj}\,k^i\,k^j\ ,
\end{equation}
\begin{equation}
\label{3.43}
F_{13}\,F_{14}+F_{23}\,F_{24}=-F^p{}_i\,F_{pj}\,k^i\,v^j+\frac{1}{2}\,F^p{}_i\,F_{pj}
\,k^i\,k^j-(F_{ij}k^i\,v^j)^2\ ,
\end{equation}
and
\begin{equation}
\label{3.44}
F_{34}=F_{ij}k^i\,v^j\ .
\end{equation}
Hence we can write
\begin{equation}
\label{3.45}
T_1=-8\,e^2F^p{}_i\,F_{pj}k^i\,v^j-16\,e^2(F_{ij}k^iv^j)^2\
.\end{equation} 
The second term $T_2$ is chosen because it involves the function $\hat c_0$ explicitly. It is defined by
\begin{equation}
\label{3.46}
T_2=\frac{1}{2}\,\hat\Delta\hat c_0-\frac{1}{2}\hat c_0\,\hat
P_0^2\left (\frac{\partial}{\partial x}\left (\hat P_0^{-2}\hat
a_{-1}\right )+\frac{\partial}{\partial y}\left (\hat P_0^{-2}\hat
b_{-1}\right )\right )-\frac{1}{2}\hat a_{-1}\,\frac{\partial\hat
c_0}{\partial x}-\frac{1}{2}\hat b_{-1}\,\frac{\partial\hat
c_0}{\partial y}\ .
\end{equation}
Explicit evaluation yields
\begin{eqnarray}
\label{3.47}
T_2&=&-6\,F_{ij}\,k^i\,v^j+\frac{1}{2}\Delta (\Delta Q_2+2\,Q_2)-3\,e^2U(u)\,\frac{\partial}{\partial x}(\log P_0)
\nonumber\\
&&-3\,e^2V(u)\,\frac{\partial}{\partial y}(\log P_0)-12\,e^2K(u)\,F_{ij}\,k^i\,v^j+11\,e^2C_{ijkl}\,k^i\,v^j\,k^k\,v^l\nonumber\\
&&-\frac{59}{3}\,e^2F^{ij}\,F_{ij}-42\,e^2(F_{ij}\,k^i\,v^j)^2-48\,e^2F^p{}_i\,F_{pj}\,v^i\,v^j\nonumber\\
&&-59\,e^2F^p{}_i\,F_{pj}\,k^i\,k^j+\frac{338}{3}\,e^2F^p{}_i\,F_{pj}\,k^i\,v^j+O_3\ .
\end{eqnarray}
The final term $T_3$ consists of the terms 
remaining in (\ref{B.6.30}). It is given by
\begin{eqnarray}
\label{3.4a}
T_3&=&3\,m\,P_0^2\left\{\frac{\partial}{\partial x}(P_0^{-2}\hat
a_0)+\frac{\partial}{\partial y}(P_0^{-2}\hat
b_0)\right\}-4\,\frac{\partial\hat f_{-1}}{\partial u}+12\,\hat
f_{-1}\,h_0+6\,m\,h_0\nonumber\\
&&-4e^2P_0^2L_2\,\frac{\partial\hat K_1}{\partial
x}-4e^2P_0^2M_2\,\frac{\partial\hat K_1}{\partial y}-2\,P_0^2\left\{\left (\frac{\partial\hat K_0}{\partial x}\right
)^2+\left (\frac{\partial\hat K_0}{\partial y}\right
)^2\right\}\nonumber\\
&&+8\,m\,P_0^2\,L_2\,\frac{\partial\hat K_0}{\partial
x}+8\,m\,P_0^2\,M_2\,\frac{\partial\hat K_0}{\partial y}
-4\,P_0^2L_2\frac{\partial\hat K_{-1}}{\partial x}-4\,P_0^2M_2\frac{\partial\hat K_{-1}}{\partial y}\nonumber\\
&&-4\,P_0^2\frac{\partial K_1}{\partial x}\,\frac{\partial\hat K_{-1}}{\partial x}-4\,P_0^2\frac{\partial K_1}{\partial y}\,\frac{\partial\hat K_{-1}}{\partial y}\nonumber\\
&&-\frac{5}{2}e^2\,P_0^2\left\{\frac{\partial}{\partial
x}(P_0^{-2}a_1)+\frac{\partial}{\partial
y}(P_0^{-2}b_1)\right\}\ .
\end{eqnarray}
When this is evaluated with the functions derived above it results in (using $h_0=a_i\,k^i$ again for convenience)
\begin{eqnarray}
\label{3.48}
T_3&=&6\,m\,h_0+2\,e^2a_i\,a^i+14\,e^2h_0^2-4\,e^2\dot h_0-4\,\dot G+12\,h_0\,G-8\,m\,e\,\dot F_{ij}\,k^i\,v^j
\nonumber\\
&&+\left (6\,m^2+\frac{5}{3}\,e^2\right )\,F^{ij}\,F_{ij}-(12\,m^2+5\,e^2)\,C_{ijkl}\,k^i\,v^j\,k^k\,v^l\nonumber\\
&&-6\,m\,X(u)\,\frac{\partial}{\partial x}(\log P_0)-6\,m\,Y(u)\,\frac{\partial}{\partial y}(\log P_0)\nonumber\\
&&+(50\,e^2-36\,m^2)\,(F_{ij}\,k^i\,v^j)^2+12\,e^2F^p{}_i\,F_{pj}\,v^i\,v^j\nonumber\\
&&-\left (36\,m^2+\frac{14}{3}\,e^2\right )\,F^p{}_i\,F_{pj}\,k^i\,v^j+(18\,m^2+5\,e^2)\,F^p{}_i\,F_{pj}\,k^i\,k^j
\nonumber\\
&&+O_3\ .
\end{eqnarray}
The approximate field equation giving us a differential equation for $Q_2$ 
is obtained from $T_1+T_2+T_3=O_3$ with $T_1, T_2, T_3$ given by (\ref{3.45}), (\ref{3.47}) and (\ref{3.48}). Making use of the unit space--like 
vector field $p^i$ introduced prior to (\ref{3.10}) we can write the result in the form
\begin{equation}
\label{3.49}
-\frac{1}{2}\Delta (\Delta Q_2+2\,Q_2)=A_0+A_1+A_2+O_3\ ,
\end{equation}
where
\begin{equation}
\label{3.50}
A_0=-4\,\dot G-\frac{16}{3}e^2F^p{}_i\,F_{pj}\,v^i\,v^j\ ,
\end{equation}
which is an $l=0$ spherical harmonic,
\begin{eqnarray}
\label{3.51}
A_1&=&6\,m\,a_i\,p^i-6\,e\,F_{ij}\,p^i\,v^j-4\,e^2h^k_i\dot a_k\,p^i-8\,m\,e\,\dot F_{ij}\,p^i\,v^j\nonumber\\
&&-8\,e^2h^k_i\,F^p{}_k\,F_{pj}\,p^i\,v^j+12\,G\,a_i\,p^i-12\,e^2K(u)\,F_{ij}\,p^i\,v^j\nonumber\\
&&-3\,(e^2U(u)+2\,m\,X(u))\,\frac{\partial}{\partial x}(\log P_0)\nonumber\\
&&-3\,(e^2V(u)+2\,m\,Y(u))\,\frac{\partial}{\partial y}(\log P_0)\ ,
\end{eqnarray}
which is an $l=1$ spherical harmonic, and 
\begin{equation}
\label{3.52}
A_2=6\,(e^2-2m^2)\,\chi _{(1)}-4\,(2\,e^2+9m^2)\,\chi
_{(2)}+18(m^2-3e^3)\,\chi _{(3)}+18e^2\chi _{(4)}\ ,
\end{equation}
which is 
an $l=2$ spherical harmonic. Any $l=2$ spherical harmonic can be written 
\begin{equation}
\label{3.53}
\chi =D_{ij}(u)\,k^i\,k^j\ ,
\end{equation}
with
\begin{equation}
\label{3.54}
D_{ij}=D_{ji}\ ,\qquad D_{ij}\,v^j=0\ ,\qquad \eta ^{ij}\,D_{ij}=0\ .
\end{equation}
In (\ref{3.52}) each $\chi _{(\alpha )}$ for $\alpha =1, 2, 3, 4$ has the form
\begin{equation}
\label{3.55}
\chi _{(\alpha )}={}_{(\alpha )}D_{ij}\,k^i\,k^j\ ,
\end{equation}
with
\begin{eqnarray}
\label{3.56}
{}_{(1)}D_{ij}&=&C_{ikjl}\,v^k\,v^l\ ,\\
{}_{(2)}D_{ij}&=&F_{ik}v^k\,F_{jl}v^l-\frac{1}{3}h_{ij}F^p{}_k\,F_{pl}\,v^k\,v^l\ ,\\
{}_{(3)}D_{ij}&=&F^p{}_k\,F_{pl}\,h^k_i\,h^l_j-\frac{1}{3}h_{ij}\,h^{kl}\,F^p{}_k\,F_{pl}\ ,\\
{}_{(4)}D_{ij}&=&a_i\,a_j-\frac{1}{3}h_{ij}\,a_k\,a^k\ .
\end{eqnarray}
The last two terms in (\ref{3.51})  can be simplified by noting that 
from (\ref{eq:P_0_in_xyru_coordinate}) and (\ref{eq:ki})  we can write
\begin{equation}
\label{3.57}
\frac{\partial}{\partial x}(\log P_0)=c_i\,k^i=c_i\,p^i\
,\qquad\frac{\partial}{\partial y}(\log P_0)=d_i\,k^i=d_i\,p^i\ ,
\end{equation}with
\begin{equation}
\label{3.58}
c_i=\left (\frac{1}{2}(v^3-v^4), 0, -\frac{1}{2}v^1, \frac{1}{2}v^1\right )\ \ {\rm and}\ \  d_i=\left (0, \frac{1}{2}(v^3-v^4), 
-\frac{1}{2}v^2, \frac{1}{2}v^2\right )\ ,
\end{equation}
covariant vectors (in coordinates $(X^i)$) defined along $r=0$ and 
everywhere orthogonal to $v^i$. Provided $A_0=O_3$, (\ref{3.49}) can be integrated 
without the introduction of directional singularities to read
\begin{equation}
\label{3.59}
\Delta Q_2+2\,Q_2=A_1+\frac{1}{3}A_2+O_3\ ,
\end{equation}
up to the addition of an arbitrary function of $u$ which makes a trivial contribution to $Q_2$. For $Q_2$ to be free of 
directional singularities we must have $A_1=O_3$ for all $p^i$ such that $p^i\,v_i=0$ and this leads to the equations of motion
\begin{eqnarray}
\label{3.60}
ma_i&=&e\,F_{ij}\,v^j+\frac{2}{3}e^2h^k_i\dot a_k+\frac{4}{3}e^2h^k_i\,F^p{}_k\,F_{pj}\,v^j\nonumber\\
&&+\frac{4}{3}m\,e\,\dot F_{ij}\,v^j-2\,G\,a_i+2\,e^2K(u)\,F_{ij}\,v^j+\hat  U(u)\,c_i+\hat V(u)\,d_i+O_3\ ,\nonumber\\
\end{eqnarray}
where we 
have written $\hat U=(e^2U(u)+2\,m\,X(u))/2=O_2$ and $\hat V=(e^2V(u)+2\,m\,Y(u))/2=O_2$. Putting $A_1=O_3$ and discarding the geometrically 
trivial homogeneous solution of (\ref{3.59}) we see that $Q_2=-A_2/12$ and this describes smooth non--trivial perturbations of the wave fronts near 
the black hole as required. $G(u)$ is given by (\ref{3.50}) with $A_0=O_3$ and thus can be written as a definite integral for which we would naturally choose 
$(-\infty , u]$ as the range of integration resulting in $-2\,G\,a_i$ in (\ref{3.60}) contributing to a `tail term'. We can write $\hat U\,c_i+\hat V\,d_i=\Omega _{ij}\,v^j$ with 
$\Omega _{ij}=-\Omega _{ji}$ vanishing except for $\Omega _{13}=\Omega _{41}=\hat U/2$ and $\Omega _{23}=\Omega _{42}=\hat V/2$. Define 
$\omega _{ij}(u)=-\omega _{ji}(u)$ by $\dot\omega _{ij}=2\,e^2K(u)\,F_{ij}+\Omega _{ij}$. If we now make the 1--parameter family of infinitesimal Lorentz transformations 
on the unit tangent vector, $v^i\rightarrow\bar v^i=v^i-(4/3)m\,e\,F^i{}_j\,v^j-(1/m)\omega ^i{} _j\,v^j$, we find that, after dropping the bar on $v$ and its derivatives 
with respect to $u$, the equations of motion take the form
\begin{equation}
\label{3.61}
m\,a_i=e\,F_{ij}\,v^j+\frac{2}{3}e^2h^k_i\dot
a_k+\frac{4}{3}e^2h^k_i\,F^p{}_k\,F_{pj}\,v^j+{\cal T}_i+O_3\ ,
\end{equation}
where 
\begin{equation}
\label{3.62}
{\cal T}_i=\frac{e}{m}\,\{\omega _k{}^j\,F_{ji}-\omega
_i{}^j\,F_{jk}-2\,G\,F_{ik}\}\,v^k=O_2\ .
\end{equation}Since $\omega _{ij}$ can be written as an integral, whose range we would 
naturally take to be $(-\infty , u]$ as in the case of $G$ above, we see that (\ref{3.62}) is a `tail term' in a set of equations of motion of the De Witt/ Brehme \cite{DB} form. Direct mathematical comparison with the equations of motion 
in \cite{DB} is highly non--trivial since the equations obtained in \cite{DB} have involved the removal of 
an infinite term, while (\ref{3.61}) does not involve such a procedure. Hence we say that (\ref{3.61}) 
fits the pattern of the equations obtained in \cite{DB}.

\setcounter{equation}{0}
\section{Discussion}\indent 
Perhaps the most significant aspect of the work presented in this paper is the fact that the equations of motion (\ref{3.61}) of a small charged black hole moving in an external electromagnetic 
and gravitational field have been derived from the vacuum Einstein--Maxwell field equations without encountering infinities. It is also important to note that the external fields 
have been introduced as a solution of the Einstein--Maxwell vacuum field equations. 

The second term on the right hand side of (\ref{3.61}) is the electromagnetic radiation reaction 4--force. The third 
term on the right hand side of (\ref{3.61}) can be written in terms of the electromagnetic energy tensor $E_{ij}(u)$ of the background space--time, calculated on $r=0$ in coordinates 
$(X^i)$. Using the background field equations on $r=0$, $R_{ij}=2\,E_{ij}$, this term can be written in terms of the background Ricci tensor components $R_{ij}$ as 
$(2/3)e^2h^k_i\,R_{kj}\,v^j$. This is the 4--force due to the external field and it differs from that derived by Hobbs \cite{HBS} by a factor of 2. The external electromagnetic 
field in \cite{HBS} is \emph{not} required to satisfy the Einstein--Maxwell vacuum field equations and so the background Ricci tensor appearing in \cite{HBS} does not have the external electromagnetic field as source. Thus in effect half of the contribution to the external 4--force is neglected.

The `tail term' in the form (\ref{3.62}) vanishes if the external electromagnetic field vanishes. It involves three 
arbitrary functions (of $u$) $K, \hat U, \hat V$ with the latter two obtained from $U, V, X, Y$ as indicated following 
(\ref{3.60}). The mathematical origin of these functions as functions of integration arising in the determination 
of $\hat K_{-1}, l_2$ and $m_2$, required for the perturbed 4--potential, and of $\hat a_0$ and $\hat b_0$ 
required for the perturbed metric tensor, is clear from this work. In addition the tail term indirectly involves the 
two vector fields $c^i$ and $d^i$ orthogonal to $v^i$. It is a topic for further study to understand physically the 
role of these arbitrary functions and of these unique space--like vectors. Had the tail term not been so explicit 
we could not identify these arbitrary functions and space--like vectors for further consideration. By comparison 
the tail term obtained in \cite{DB} is an integral whose integrand is expressed in terms of functions appearing in the 
Hadamard form of the Green function of the vector wave equation. Very little is known explicitly about the form of 
such functions. The De Witt-Brehme tail term has proved quite intractable to analyse due to the fact that the Green 
function is not known in closed form. This makes a comparison with our result extremely difficult. Such a study would surely merit a paper independently of ours.

The original calculations of Dirac \cite{D} were carried out in Minkowskian space--time. Our results do not specialise to those of Dirac since our background 
space--time is an Einstein--Maxwell space--time and if it is flat then the external electromagnetic field vanishes. In this case (\ref{3.61}) gives $m\,a_i=O_2$ and when 
this is substituted into the surviving radiation reaction 4--force the latter is absorbed into the $O_3$--error. Hence we obtain geodesic motion (and in particular no ``run--away'' motion) at this level of approximation 
in this theory.
\setcounter{equation}{0}
\section*{Acknowledgement}\indent
We are grateful to Takashi Fukumoto for carrying out helpful
algebraic computer calculations in the early stages of this work.
P.A.H. thanks the JSPS for a Long-Term Invitation Fellowship. This work is partly 
supported by a Grant-in-Aid for the G-COE Program ``Weaving Science Web Beyond Particle--Matter 
Hierarchy'' in Tohoku University, funded by the Ministry of Education, Culture, Sports, Science 
and Technology (MEXT) of Japan. Y.I. is in part supported by a Grant-in-Aid for Young Scientists 
(B)(20740118) from MEXT.

\appendix
\section{Useful Formulas from Minkowskian Geometry} 

We record here some useful equations which hold exactly in Minkowskian
space--time and are also important in the neighbourhood of the
world line $r=0$ in the background space--time in this paper. The
transformation (\ref{eq:Xi_to_xi}) can in principle be inverted giving
$(x, y, r, u)$ as functions of $(X^i)$ which means that we can
consider $(x, y, r, u)$ as scalar fields on Minkowskian
space--time. Their derivatives with respect to $(X^i)$, denoted by
a comma, are obtained by first differentiating (\ref{eq:Xi_to_xi}) to
arrive at
\begin{equation}
\label{A1}
\delta
^i_j=(v^i-r\,h_0\,k^i)\,u_{,j}+k^i\,r_{,j}+r\,\frac{\partial
k^i}{\partial x}\,x_{,j}+r\,\frac{\partial k^i}{\partial
y}\,y_{,j}\ .
\end{equation}
Multiplying this by $k_i$ gives immediately
\begin{equation}
\label{A2}
k_j=u_{,j}\ .
\end{equation} 
Now multiplying (\ref{A1}) by $v_j$
yields
\begin{equation}
\label{A3}
r_{,j}=v_j-(1-r\,h_0)\,k_j\ .
\end{equation}

Differentiating (\ref{eq:Xi_to_xi}) with respect to $X^j$ and using (\ref{A2}) 
and (\ref{A3}) results in the alternative form for (\ref{A1}):
\begin{equation}\label{A1'}
k^i{}_{,j}=\frac{1}{r}\left (\delta ^i_j-v^i\,k_j-k^i\,v_j+(1-r\,h_0)\,k^i\,k_j\right )\ ,\end{equation}
with $h_0=a_i\,k^i$. From this it follows that
\begin{equation}\label{A1''}
\frac{\partial k^i}{\partial r}=k^i{}_{,j}\,k^j=0\ ,
\qquad\frac{\partial k^i}{\partial u}=k^i{}_{,j}\,v^j=-h_0\,k^i\ .\end{equation}
Writing $h_0=\partial (\log P_0)/\partial u$ for some function $P_0$ independent of $r$ we 
see from the latter two equations that $k^i=P_0^{-1}\,\zeta ^i$ with $\zeta ^i$ null and 
independent of $r$ and $u$. Thus $\zeta ^i$ can be parametrized by two real parameters. We have 
chosen the parameters $x, y$ as given in (\ref{eq:ki}) and then $P_0=\zeta _i\,v^i$ in (\ref{eq:P_0_in_xyru_coordinate}) 
is a consequence of this choice. A more complete discussion of this construction can be found in 
\cite{Mo} and \cite{HoEl}.
 
We can now simplify (\ref{A1}) to read
\begin{equation}
\label{A4}
\delta ^i_j=v^i\,k_j+k^i\,v_j-k^i\,k_j+r\,\frac{\partial
k^i}{\partial x}\,x_{,j}+r\,\frac{\partial k^i}{\partial
y}\,y_{,j}\ .
\end{equation}
Multiplying (\ref{A4}) by $\partial
k_i/\partial x$ and also by $\partial k_i/\partial y$ provides the
remaining equations needed:
\begin{equation}
\label{A5}
x_{,j}=-\frac{1}{r}P_0^2\,\frac{\partial k_j}{\partial x}\ ,\qquad
y_{,j}=-\frac{1}{r}P_0^2\,\frac{\partial k_j}{\partial y}\
.\end{equation}
Substituting (\ref{A5}) into (\ref{A4}) results in
the Minkowskian metric tensor being written in the form
\begin{equation}
\label{A6}
\eta ^{ij}=-P_0^2\left (\frac{\partial k^i}{\partial
x}\,\frac{\partial k^j}{\partial x}+\frac{\partial k^i}{\partial
y}\,\frac{\partial k^j}{\partial y}\right
)+k^i\,v^j+k^j\,v^i-k^i\,k^j\ ,
\end{equation}
and much use is made of this 
relation in the calculations behind this paper. Equivalently the
Minkowskian line--element takes the form
\begin{equation}
\label{A7}
ds^2=\eta
_{ij}\,dX^i\,dX^j=-r^2P_0^{-2}(dx^2+dy^2)+2\,du\,dr+(1-2\,h_0\,r)\,du^2\
.\end{equation}
This suggests we introduce basis 1--forms (defining
a half null basis)
\begin{equation}
\label{A8}
\vartheta ^1=r\,P_0^{-1}dx\ ,\ \vartheta ^2=r\,P_0^{-1}dy\ ,\
\vartheta ^3=dr+\frac{1}{2}(1-2\,h_0\,r)\,du\ ,\ \vartheta ^4=du\
.\end{equation}Using (\ref{A2}), (\ref{A3}) and (\ref{A5}) we can
express these in terms of the rectangular Cartesian coordinates
and time as
\begin{eqnarray}
\vartheta ^1&=&-P_0\,\frac{\partial k_i}{\partial x}\,dX^i=-\vartheta
 _1\ ,
\label{A9}
\\
\vartheta ^2&=&-P_0\,\frac{\partial k_i}{\partial y}\,dX^i=-\vartheta
 _2\ ,
\label{A10}
\\
\vartheta ^3&=&(v_i-\frac{1}{2}k_i)\,dX^i=\vartheta _4\ ,
\label{A11}
\\
\vartheta ^4&=&k_i\,dX^i=\vartheta _3\ .
\label{A12}
\end{eqnarray}

It is helpful to have available the second partial derivatives of
$k^i$ with respect to $x$ and $y$ expressed on the basis
consisting of the vectors $v^i,\ k^i,\ \partial k^i/\partial x$
and $\partial k^i/\partial y$. These formulas are:
\begin{eqnarray}
%\label{A10}
\frac{\partial ^2k^i}{\partial
x^2}&=&P_0^{-2}(v^i-k^i)-\frac{\partial}{\partial x}(\log
P_0)\,\frac{\partial k^i}{\partial x}+\frac{\partial}{\partial
y}(\log P_0)\,\frac{\partial k^i}{\partial y}\ ,
\label{A13}
\\
\frac{\partial ^2k^i}{\partial
y^2}&=&P_0^{-2}(v^i-k^i)+\frac{\partial}{\partial x}(\log
P_0)\,\frac{\partial k^i}{\partial x}-\frac{\partial}{\partial
y}(\log P_0)\,\frac{\partial k^i}{\partial y}\ ,
\label{A14}
\\
\frac{\partial ^2k^i}{\partial x\partial
y}&=&-\frac{\partial}{\partial y}(\log P_0)\,\frac{\partial
k^i}{\partial x}-\frac{\partial}{\partial x}(\log
P_0)\,\frac{\partial k^i}{\partial y}\ .
\label{A15}
\end{eqnarray}

\section{Perturbed Field Equations} \setcounter{equation}{0}
Writing, for the perturbed space--time described in section 3, ${\cal M}^a=\hat F^{ab}{}_{|b}$ and 
${\cal E}_{ab}=\hat R_{ab}-2\,\hat E_{ab}$ the leading terms and the errors we tolerate in these 
expressions (and thus the extent to which we satisfy the vacuum Einstein--Maxwell field equations) are 
given here. If the equations of motion are required with greater accuracy then the 
field equations have to be solved with greater accuracy resulting in smaller errors in the coefficients 
of these powers of $r$.
\begin{eqnarray}
{\cal M}^A&=&O_3\times r^{-2}+O_2\times r^{-1}+O_1+O(r)\ ,\ (A=1, 2), 
\label{B.1.1}
\\
{\cal M}^3&=&O_3\times r^{-1}+O_1+O(r)\ ,
\label{B.1.2}\\
{\cal M}^4&=&O_2\times r+O(r^2)\ ,
\label{B.1.3}
\end{eqnarray}and 
\begin{eqnarray}
{\cal E}_{AA}&=&O_3\times r^{-4}+O_3\times r^{-2}
+O_2\times r^{-1}+O(r^0)\ ,
\label{B.2.1}
\\
{\cal E}_{11}-{\cal E}_{22}&=&O_3\times r^{-2}+O_2\times r^{-1}+O_1+O(r)
\ , 
\label{B.2.2}
\\
{\cal E}_{12}&=&O_3\times r^{-2}+O_2\times r^{-1}+O_1+O(r)\ ,
\label{B.2.3}
\\
{\cal E}_{A3}&=&O_3\times r^{-2}+O_1\times r^{-1}+O_1+O(r)\ ,
\label{B.2.4}
\\
{\cal E}_{A4}&=&O_2\times r^{-3}+O_2\times r^{-2}+O_2\times r^{-1}
+O(r^0)\ ,
\label{B.2.5}
\\
{\cal E}_{33}&=&O_1+O(r)\ ,
\label{B.2.6}
\\
{\cal E}_{34}&=&O_4\times r^{-4}+O_3\times r^{-3}+O_2\times r^{-2}\nonumber\\
&&+O_2\times r^{-1}+O(r^0)\ ,
\label{B.2.7}
\\
{\cal E}_{44}&=&O_3\times r^{-3}+O_3\times r^{-2}+O(r^{-1})\ .
\label{B.2.8}
\end{eqnarray}
The coefficients of the various powers of $r$ in these 
expressions are calculated by substituting into ${\cal M}^a$ and ${\cal
E}_{ab}$ the metric tensor, given 
via the line--element (\ref{eq:metric}) 
with the expansions (\ref{3.1.1})--(\ref{3.1.6}),  
and the potential 1--form (\ref{eq:maxwell_potential}), with the
expansions (\ref{3.2.1})--(\ref{3.2.3}). Writing 
(\ref{B.1.1})--(\ref{B.2.8}) 
in the form
\begin{equation}
{\cal M}^a=\sum {}_{(n)}{\cal M}^a\,r^n\ ,\qquad {\cal E}_{ab}=\sum
{}_{(n)}{\cal E}_{ab}\,r^n\ ,
\label{B.3}
\end{equation}
the coefficients ${}_{(n)}{\cal M}^a$ and ${}_{(n)}{\cal E}_{ab}$ required to establish (\ref{B.1.1})--(\ref{B.2.8})  are:
\begin{eqnarray}
{}_{(-2)}{\cal M}^1&=&\frac{\partial}{\partial y}\left\{\hat P_0^2\left (\frac{\partial\hat M_0}{\partial x}-
\frac{\partial\hat L_0}{\partial y}\right )\right\}+2e^2\hat L_2-
\,e \, \hat a_{-1}\,\hat P_0^{-2}\nonumber\\
&&+\frac{\partial\hat K_{-1}}{\partial x}+O_3\ ,
\label{B.4.0}
\\
{}_{(-1)}{\cal M}^1&=&O_3\ ,
\label{B.4.1}
\\
{}_{(0)}{\cal M}^1&=&6\,m\,L_3+O_2=O_1\ ,
\label{B.4.2}
\\
{}_{(-2)}{\cal M}^2&=&\frac{\partial}{\partial x}\left\{\hat P_0^2\left (\frac{\partial\hat M_0}{\partial x}-
\frac{\partial\hat L_0}{\partial y}\right )\right\}-2e^2\hat M_2+
\, e \, \hat b_{-1}\hat P_0^{-2}\nonumber\\
&&-\frac{\partial\hat K_{-1}}{\partial y}+O_3\ ,
\label{B.4.3}
\\
{}_{(-1)}{\cal M}^2&=&O_3\ ,
\label{B.4.4}
\\
{}_{(0)}{\cal M}^2&=&-6\,m\,M_3+O_2=O_1\ ,
\label{B.4.5}
\\
{}_{(-1)}{\cal M}^3&=&O_2\ ,
\label{B.4.6}
\\
{}_{(0)}{\cal M}^3&=&-4mP_0^{-2}F_{ij}k^i\,v^j+O_2=O_1\ ,
\label{B.4.7}
\\
{}_{(1)}{\cal M}^4&=&-2\hat P_0^{-2}\hat K_1+4\hat P_0^{-2}\,(\hat a_{-1}\hat L_2+\hat b_{-1}\hat M_2
-e\,\hat q_2)\nonumber\\
&&+2\left (\frac{\partial\hat L_2}{\partial x}
+\frac{\partial\hat M_2}{\partial y}\right )+O_2\ .
\label{B.4.8}
\end{eqnarray}
Writing
\begin{equation}
\label{B.5}
\hat\Delta =\hat P_0^2\left (\frac{\partial ^2}{\partial
		       x^2}+\frac{\partial ^2}{\partial y^2}\right )\ ,
\end{equation}
we find that 
\begin{eqnarray}
{}_{(-4)}{\cal E}_{AA}&=&O_3\ ,
\label{B.6.1}
\\
{}_{(-3)}{\cal E}_{AA}&=&O_3\ ,
\label{B.6.2}
\\
{}_{(-2)}{\cal E}_{AA}&=&2\,\hat\Delta\log\hat P_0-4\,\hat P_0^4\left(\frac{\partial\hat M_0}
{\partial x}-\frac{\partial\hat L_0}{\partial y}\right )\,\left (\frac{\partial\hat M_2}{\partial x}
-\frac{\partial\hat L_2}{\partial y}\right )\nonumber\\
&&-3\hat P_0^2\left (\frac{\partial}{\partial x}(\hat P_0^{-2}\hat a_{-1})
+\frac{\partial}{\partial y}(\hat P_0^{-2}\hat b_{-1})\right )\nonumber\\
&&+4\,e\hat K_1+4\,\hat K_1\,\hat K_{-1}-8\,e(\hat a_{-1}\hat L_2+\hat b_{-1}\hat M_2)\nonumber\\
&&-2\,\hat c_0+12\,e^2\hat q_2
-\frac{1}{2}\hat P_0^{-2}(\hat a_{-1}^2+\hat b_{-1}^2)+O_3\ ,
\label{B.6.3}
\\
{}_{(-1)}{\cal E}_{AA}&=&-4\,\hat c_1-8\,\frac{\partial\log\hat P_0}{\partial u}+8\,e\,\hat K_2
-32\,m\,\hat q_2\nonumber\\
&&-4\,\hat P_0^2\left (\frac{\partial}{\partial x}(\hat P_0^{-2}\hat a_0)
+\frac{\partial}{\partial y}(\hat P_0^{-2}\hat b_0)\right )+O_2\ ,
\label{B.6.4}
\\
{}_{(-2)}{\cal E}_{11}-{}_{(-2)}{\cal E}_{22}&=&\frac{\partial\hat b_{-1}}{\partial y}-\frac{\partial\hat a_{-1}}{\partial x}
-4\,e^2\hat\alpha _2-\frac{1}{2}\hat P_0^{-2}(\hat a_{-1}^2+\hat b_{-1}^2)\nonumber\\
&&-8\,e^2\hat P_0^2(\hat L_2^2-\hat M_2^2)-8\,\hat P_0^2\hat L_2\,\frac{\partial\hat K_{-1}}{\partial x}+
8\,\hat P_0^2\hat M_2\,\frac{\partial\hat K_{-1}}{\partial y}\nonumber\\
&&+O_3\ ,
\label{B.6.5}
\\
{}_{(-1)}{\cal E}_{11}-{}_{(-1)}{\cal E}_{22}&=&16\,m\,\hat P_0^2(\hat L_2^2-\hat M_2^2)-8\,\hat P_0^2\hat L_2\,\frac{\partial\hat K_0}{\partial x}+8\,\hat P_0^2\hat M_2\,\frac{\partial\hat K_0}{\partial y}
\nonumber\\&&+16\,m\,\hat\alpha _2-2\,\frac{\partial\hat a_0}{\partial
x}+2\,\frac{\partial\hat b_0}{\partial y}+O_2\ ,
\label{B.6.6}
\\
{}_{(0)}{\cal E}_{11}-{}_{(0)}{\cal E}_{22}&=&-8\hat c_0\,\hat P_0^2(\hat L_2^2-\hat M_2^2)
-8\,\hat P_0^2\hat L_2\,\frac{\partial{\hat K_1}}{\partial x}+8\,\hat P_0^2\hat M_2\,\frac{\partial{\hat K_1}}{\partial y}
 \nonumber\\
&&-12\,\hat\alpha_2\,\hat c_0 -3\frac{\partial{\hat
 a_1}}{\partial x}+3\frac{\partial{\hat b_1}}{\partial y}
 + O_1=O_1\ ,
\label{B.6.7}
\\
{}_{(-2)}{\cal E}_{12}&=&-\frac{1}{2}\frac{\partial\hat a_{-1}}{\partial y}-\frac{1}{2}\frac{\partial\hat b_{-1}}{\partial x}
-4\,e^2\hat\beta _2-\frac{1}{2}\hat P_0^{-2}\hat a_{-1}\,\hat b_{-1}\nonumber\\
&&-8\,e^2\hat P_0^2\hat L_2\hat M_2-4\,\hat P_0^2\hat L_2\,\frac{\partial\hat K_{-1}}{\partial y}-
4\,\hat P_0^2\hat M_2\,\frac{\partial\hat K_{-1}}{\partial x}\nonumber\\
&&+O_3\ ,
\label{B.6.8}
\\
{}_{(-1)}{\cal E}_{12}&=&16\,m\,\hat P_0^2\hat L_2\,\hat M_2-4\,\hat P_0^2\hat L_2\,
\frac{\partial\hat K_0}{\partial y}-4\,\hat P_0^2\hat M_2\,\frac{\partial\hat K_0}{\partial x}
\nonumber\\&&+8\,m\,\hat\beta _2-\frac{\partial\hat a_0}{\partial
y}-\frac{\partial\hat b_0}{\partial x}+O_2\ ,
\label{B.6.9}
\\
{}_{(0)}{\cal E}_{12}&=&-8\hat c_0\,\hat P_0^2\hat L_2\,\hat M_2
-4\,\hat P_0^2\hat M_2\,\frac{\partial{\hat K_1}}{\partial x}-4\,\hat P_0^2\hat L_2\,\frac{\partial{\hat K_1}}{\partial y}
 \nonumber\\&&-6\,\hat\beta_2\,\hat c_0-\frac{3}{2}\frac{\partial{\hat
 a_1}}{\partial y}-\frac{3}{2}\frac{\partial{\hat b_1}}{\partial x}
 + O_1=O_1\ ,
\label{B.6.10}
\\
{}_{(-2)}{\cal E}_{13}&=&4\,\hat M_2\,\hat P_0^3\left(
\frac{\partial{\hat L_0}}{\partial y}-\frac{\partial{\hat M_0}}{\partial x}\right)+4\,e\,\hat L_2
\hat P_0+\hat P_0^{-1}\hat a_{-1}\nonumber\\
&&+4\,\hat K_{-1}\,\hat L_2\,\hat P_0+O_3\ ,
\label{B.6.11}
\\
{}_{(-1)}{\cal E}_{13}&=&6\,e\,P_0\,L_3+O_2=O_1\ ,
\label{B.6.12}
\\
{}_{(0)}{\cal E}_{13}&=&-4\,\hat K_1\,\hat L_2\,\hat P_0+2\,\hat P_0^{3}\left (\frac{\partial}{\partial x}(\hat P_0
^{-2}\hat\alpha _2)+\frac{\partial}{\partial y}(\hat P_0^{-2}\hat\beta _2)\right )\nonumber\\
&&+2\,\hat P_0\,\frac{\partial\hat q_2}{\partial x}-2\,\hat P_0^{-1}\hat a_1+4\,\hat P_0^3\hat M_2\,\left (\frac{\partial\hat L_2}{\partial y}
-\frac{\partial\hat M_2}{\partial x}\right )+O_1\nonumber\\
&&=O_1\ ,
\label{B.6.13}
\\
{}_{(-3)}{\cal E}_{14}&=&4\,m\,e\,P_0\,F_{ij}\frac{\partial
 k^i}{\partial x}\,v^j+O_3=O_2\ ,
\label{B.6.14}
\\
{}_{(-2)}{\cal E}_{14}&=&2\,e\,\hat c_0\hat L_2\,\hat P_0+2\,e\,\hat P_0\frac{\partial\hat K_1}
{\partial x}+\frac{1}{2}\,\hat P_0^{-1}\hat a_{-1}\,\hat c_0\nonumber\\
&&-\frac{1}{2}\hat P_0\frac{\partial}{\partial y}\left (\hat P_0^2\left\{\frac{\partial}
{\partial y}(\hat P_0^{-2}\hat a_{-1})-\frac{\partial}
{\partial x}(\hat P_0^{-2}\hat b_{-1})\right\}\right )\nonumber\\
&&-\hat P_0^{-1}\hat a_{-1}\,\hat\Delta\log\hat P_0+O_2=O_2\ ,
\label{B.6.15}
\\
{}_{(-1)}{\cal E}_{14}&=&O_2\ ,
\label{B.6.16}
\\
{}_{(-2)}{\cal E}_{23}&=&-4\,\hat L_2\,\hat P_0^3\left(
\frac{\partial{\hat L_0}}{\partial y}-\frac{\partial{\hat M_0}}{\partial x}\right)+4\,e\,\hat M_2
\hat P_0+\hat P_0^{-1}\hat b_{-1}\nonumber\\
&&+4\,\hat K_{-1}\,\hat M_2\,\hat P_0+O_3\ ,
\label{B.6.17}
\\
{}_{(-1)}{\cal E}_{23}&=&6\,e\,P_0\,M_3+O_2=O_1\ ,
\label{B.6.18}
\\
{}_{(0)}{\cal E}_{23}&=&-4\,\hat K_1\,\hat M_2\,\hat P_0+2\,\hat P_0^{3}\left (-\frac{\partial}{\partial y}(\hat P_0
^{-2}\hat\alpha _2)+\frac{\partial}{\partial x}(\hat P_0^{-2}\hat\beta _2)\right )\nonumber\\
&&+2\,\hat P_0\,\frac{\partial\hat q_2}{\partial y}-2\,\hat P_0^{-1}\hat b_1-4\,\hat P_0^3\hat L_2\,\left (\frac{\partial\hat L_2}{\partial y}
-\frac{\partial\hat M_2}{\partial x}\right )+O_1\nonumber\\
&&=O_1\ ,
\label{B.6.19}
\\
{}_{(-3)}{\cal E}_{24}&=&4\,m\,e\,P_0\,F_{ij}\frac{\partial
 k^i}{\partial y}\,v^j+O_3=O_2\ ,
\label{B.6.20}
\\
{}_{(-2)}{\cal E}_{24}&=&2\,e\,\hat c_0\hat M_2\,\hat P_0+2\,e\,\hat P_0\frac{\partial\hat K_1}
{\partial y}+\frac{1}{2}\,\hat P_0^{-1}\hat b_{-1}\,\hat c_0\nonumber\\
&&+\frac{1}{2}\hat P_0\frac{\partial}{\partial x}\left (\hat P_0^2\frac{\partial}
{\partial y}(\hat P_0^{-2}\hat a_{-1})-\hat P_0^2\frac{\partial}
{\partial x}(\hat P_0^{-2}\hat b_{-1})\right )\nonumber\\
&&-\hat P_0^{-1}\hat b_{-1}\,\hat\Delta\log\hat P_0+O_2=O_2\ ,
\label{B.6.21}\\
{}_{(-1)}{\cal E}_{24}&=&O_2\ ,
\label{B.6.22}
\\
{}_{(0)}{\cal E}_{33}&=&-8\,\hat P_0^2(\hat L_2^2+\hat M_2^2)+12\,\hat
 q_2+O_3\ ,
\label{B.6.23}
\\
{}_{(-4)}{\cal E}_{34}&=&O_4\ ,
\label{B.6.24}
\\
{}_{(-3)}{\cal E}_{34}&=&O_3\ ,
\label{B.6.25}
\\
{}_{(-2)}{\cal E}_{34}&=&2\,e\,\hat K_1+\frac{1}{2}\hat P_0^2\left (\frac{\partial}{\partial x}
(\hat P_0^{-2}\hat a_{-1})+\frac{\partial}{\partial y}(\hat P_0^{-2}\hat b_{-1})\right )\nonumber\\
&&+O_2\ ,
\label{B.6.26}
\\
{}_{(-1)}{\cal E}_{34}&=&\hat c_1+2\,\hat P_0^{-1}\frac{\partial\hat P_0}{\partial u}+4\,e\,\hat K_2
+8\,m\,\hat q_2\nonumber\\
&&+\hat P_0^2\left (\frac{\partial}{\partial x}(\hat P_0^{-2}\hat a_0)+
\frac{\partial}{\partial y}(\hat P_0^{-2}\hat b_0)\right )+O_2\ ,
\label{B.6.27}
\\
{}_{(-4)}{\cal E}_{44}&=&O_3\ ,
\label{B.6.28}
\\
{}_{(-3)}{\cal E}_{44}&=&2\,m\,\hat P_0^2\left (\frac{\partial}{\partial x}(\hat P_0^{-2}\hat a_{-1})+
\frac{\partial}{\partial y}(\hat P_0^{-2}\hat b_{-1})\right )\nonumber\\
&&-4\,e^2\hat P_0\,\frac{\partial\hat P_0}{\partial u}-2\,\hat\Delta\hat
 f_{-1}+O_3\,
\label{B.6.29}
\\
{}_{(-2)}{\cal E}_{44}&=&T_1+T_2+T_3+O_3\ .
\label{B.6.30}
\end{eqnarray}
In the final equation here the terms $T_1,\ 
T_2,\ T_3$ are given in the text by (\ref{3.36}), (\ref{3.46}) and
(\ref{3.4a}) respectively.


\begin{thebibliography}{99}
\bibitem{DB} B. S. De Witt and R. W. Brehme, Ann. Phys. NY {\bf 9}, 220 (1960).
\bibitem{D} P. A. M. Dirac, Proc. R. Soc. A{\bf 167}, 148 (1932).
\bibitem{B} R. Beig, Acta Phys. Asut. {\bf 38}, 300 (1973).
\bibitem{BV} A. O. Barut and D. Villaroel, J. Phys. A: Math. Gen. {\bf 8}, L1537 (1975).
\bibitem{M} Y. Mino, M. Sasaki and T. Tanaka, Phys. Rev. D{\bf
55}, 3457 (1997).
\bibitem{Q} T. C. Quinn and R. M. Wald, Phys. Rev. D{\bf 56}, 3381
(1997).
\bibitem{DW} S. Detweiler and B. F. Whiting, Phys. Rev. D{\bf 67},
024025 (2003).
\bibitem{P} E. Poisson, Class. and Quantum Grav. {\bf 21}, R153
(2004).
\bibitem{P1} E. Poisson, Living Rev. Rel. {\bf 7}, 6 (2004).
\bibitem{GW} S. E. Gralla and R. M. Wald, Class. and Quantum Grav. {\bf 25}, 205009 (2008).
\bibitem{Bar} L. Barack, Y. Mino, H. Nakano, A. Ori and M. Sasaki, Phys. Rev.
Lett. {\bf 88}, 091101 (2002).
\bibitem{Bar1} L. Barack and A. Ori, Phys. Rev. D{\bf 66}, 084022
(2002).
\bibitem{Bar2} L. Barack and A. Ori, Phys. Rev. D{\bf 67}, 024029
(2003).
\bibitem{Bar3} L. Barack and A. Ori, Phys. Rev. Lett. {\bf 90},
111101 (2002).
\bibitem{MNS} Y. Mino, H. Nakano and M. Sasaki, Prog. Theor. Phys.
{\bf 108}, 1039 (2003).
\bibitem{DMW} S. Detweiler, E. Messaritaki and B. F. Whiting,
Phys. Rev. D{\bf 67}, 104016 (2003).
\bibitem{Bar4} L. Barack and C. O. Lousto, Phys. Rev. D{\bf 66},
061502 (2002).
\bibitem{PfP} M. J. Pfenning and E. Poisson, Phys. Rev. D{\bf 65},
084001 (2002).
\bibitem{HR} P. A. Hogan and I. Robinson, Found. Phys. {\bf 15},
617 (1985).
\bibitem{NP} E. T. Newman and R. Posadas, Phys. Rev. {\bf 187},
1784 (1969).
\bibitem{RR} I. Robinson and J. R. Robinson, in \emph{General
Relativity: papers in honour of J. L. Synge}, ed. L.
O'Raifeartaigh (Clarendon Press, Oxford 1972), p.151.
\bibitem{HI} P. A. Hogan and M. Imaeda, J. Phys. A: Math. Gen.
{\bf 12}, 1061 (1979).
\bibitem{HT} P. A. Hogan and A. Trautman, in \emph{Gravitation and
Geometry}, edited by W. Rindler and A. Trautman (Bibliopolis,
Naples, 1987), p. 215.
\bibitem{FH} T. Futamase and P. A. Hogan, J. Math. Phys. {\bf 34},
154 (1993).
\bibitem{S} R. K. Sachs, Proc. R. Soc. London, Ser. A{\bf 265}, 463
(1962).
\bibitem{NU} E. T. Newman and T. W. J. Unti, J. Math. Phys. {\bf
3}, 891 (1962).
\bibitem{RT1} I. Robinson and A. Trautman, Phys. Rev. Lett. {\bf
4}, 431 (1960).
\bibitem{RT2} I. Robinson and A. Trautman, Proc. R. Soc. London,
Ser. A{\bf 270}, 463 (1962).
\bibitem{EF} E. Fermi, Atti Accad. Naz. Lincei Rend. Cl. Sci. Fis.
Mat.Nat. {\bf 31}, 21 and 51 (1922).
\bibitem{OR1} L. O'Raifeartaigh, Proc. R. Irish Acad. Ser. A{\bf
59}, 15 (1958).
\bibitem{OR2} L. O'Raifeartaigh, Proc. R. Irish Acad. Ser. A{\bf
62}, 63 (1962).
\bibitem{NU1} E. T. Newman and T. W. J. Unti, J. Math. Phys. {\bf 4}, 
1467 (1969).
\bibitem{Mo} T. Molenda, ``Contributed Papers, vol.1, 10th. International 
Conference on General Relativity and Gravitation 1983''(B. Bertotti, F. de Felice 
and A. Pascolini, Eds.), p.97.
\bibitem{HoEl} P. A. Hogan and G. F. R. Ellis, Ann. Phys. NY {\bf 195}, 293 (1989).
\bibitem{HBS} J. M. Hobbs, Ann. Phys. NY {\bf 47}, 141 (1968).
\end{thebibliography}
\end{document}